\documentclass[12pt]{article}        

\usepackage{color,graphicx}
\usepackage{latexsym,amssymb,amsmath}
\numberwithin{equation}{section}

\newtheorem{defi}{Definition}[section]
\newtheorem{rem}{Remark}[section]

\newcommand{\bsquare}{\hbox{\rule{6pt}{6pt}}}

\newtheorem{theorem}{Theorem}

\newtheorem{lmm}{Lemma}[section]

\begin{document}

\title{Direct construction of scalar quantum fields by L{\'e}vy fields\\
-
non trivial  exact Wightman fields in 
a wider field with a relaxed   
 G{\aa}rding-Wightman Axioms-
}


\author{
{{Sergio
 \textsc{Albeverio}}}
          \footnote{Inst. Angewandte Mathematik, 
and HCM, Univ. Bonn, Germany, 
}
\quad 
{{Suji \textsc{Kawasaki}}}
\footnote{Dept. Math., Iwate Univ., Japan,}, 
 \quad
 {{Yumi \textsc{Yahagi}}}
\footnote{Dept. Math., Tokyo Gakugei Univ., Japan,}, 
\\
{{Minoru \textsc{W. Yoshida}}} \footnote{Dept. Information Systems Kanagawa Univ., Yokohama, Japan,  
 email:\, washizuminoru@hotmail.com }
 }


\maketitle


{\textcolor{blue}{
{\large{Presentation version for the conferences on Linaeus 2024 and QBIC 2025.
}}
}}

\begin{abstract} 
{\textcolor{blue}{
This paper introduces partial results, in the current situation, 
of ongoing considerations 
corresponding to the above title. 
 A construction on exact relativistic quantum field model with the space time dimension $d \in {\mathbb N}$,  including the case where $d \geq 4$, is going to 
 be discussed.}}

Firstly, 
Hermitian scalar quantum fields
 $<{\cal H}, U, \psi, D>$, 
within a relaxed framework of the G{\aa}rding-Wightman Axioms, is constructed 
 by making use of the stochastic calculus arguments 
 with respect to the
 {\it{stationary additive random fields }} on ${\mathbb R}^d$, i.e., 
  the {\it{L{\'e}vy random fields}} on ${\mathbb R}^d$.
The first constructed $<{\cal H}, U, \psi, D>$, here, 
 satisfy all the requirements of the the G{\aa}rding-Wightman Axioms, except that the field operators $\psi (f)$ with $f \in  {\cal S}({\mathbb R}^d \to {\mathbb R})$ are symmetric operators on the physical Hilbert space ${\cal H}$, which situation is denoted here as 
{\it{a relaxed framework}} of the G{\aa}rding-Wightman Axioms.

Secondly, by taking the adequate subspaces of ${\cal H}$, 
non trivial  exact Wightman quantum fields,  
which satisfy all the requirements of the G{\aa}rding-Wightman Axioms, 
are constructed actually.

\medskip

\noindent
{\bf{keywords}}:  {\footnotesize{Axiomatic quantum field theory, G{\aa}rding-Wightman axioms, Bochner-Minlos theorem, L{\'e}vy fields on ${\mathbb R}^d$.}}
\medskip

\end{abstract}

\section{Introduction}
{\textcolor{blue}{
This paper introduces partial results, in the current situation, 
of ongoing considerations 
corresponding to the given title.
A construction on exact relativistic quantum field model with the space time dimension $d \in {\mathbb N}$, including the case where $d \geq 4$,
is going to be discussed.}}

In short, 
firstly, 
Hermitian scalar quantum fields
 $<{\cal H}, U, \psi, D>$, 
within a relaxed framework of the G{\aa}rding-Wightman Axioms, is constructed 
 by making use of the stochastic calculus arguments with respect to the
 {\it{stationary additive random fields }} on ${\mathbb R}^d$, i.e., 
  the {\it{L{\'e}vy random fields}} on ${\mathbb R}^d$.
The first constructed $<{\cal H}, U, \psi, D>$, here, 
 satisfy all the requirements of the the G{\aa}rding-Wightman Axioms, except that the field operators $\psi (f)$ with $f \in  {\cal S}({\mathbb R}^d \to {\mathbb R})$ are symmetric operators on the physical Hilbert space ${\cal H}$, which situation is denoted here as 
{\it{a relaxed framework}} of the G{\aa}rding-Wightman Axioms.
Secondly, by taking the adequate subspaces of ${\cal H}$, 
non trivial  exact Wightman quantum fields,  
which satisfy all the requirements of the G{\aa}rding-Wightman Axioms, 
are constructed actually.

A little in detail, 
firstly, 
by making use of the {\it{real}} L{\'e}vy random fields on ${{\mathbb R}^d}$ (
or the centered {\it{real}} Gaussian random field on ${{\mathbb R}^d}$, as a benchmark of the models),
a relaxed quantum field $<{\cal H}, U, \psi, D>$, of which  Hilbert space ${\cal H}$  is sufficiently wide, 
is constructed. 
Here the  
(multiplicative) field operators $\psi$ 
are precisely as field operators $\psi^{+} (f)$ or $\psi^{-} (f)$ for $f \in  {\cal S}({\mathbb C}^d \to {\mathbb R})$ ($\psi^{+}$ and $\psi^{-}$ are complex conjugate each other, in some sense). 
By the first construction $<{\cal H}, U, \psi, D>$ 
 satisfies all the requirements of the the G{\aa}rding-Wightman Axioms, except 
 the requirement such 
that the field operators $\psi (f)$ with $f \in  {\cal S}({\mathbb R}^d \to {\mathbb R})$ are symmetric operators on the physical Hilbert space ${\cal H}$.

Next,  for each {\it{real valued}} $f_{\mathbb R} \in  {\cal S}({\mathbb R}^d \to {\mathbb R})$, the combination of the (multiplicative) operators $\psi^{+} (f_{\mathbb R}) + \psi^{-} (f_{\mathbb R})$, denoted by $\psi^{\cos} (f_{\mathbb R})$ (
 $\psi^{\cos} \, : 
{\cal S}({\mathbb R}^d \to {\mathbb C}) \ni 
f_{\mathbb C} 
\longmapsto
 \psi^{\cos} (f_{\mathbb C})$
 is an (multiplicative) operator valued {\it{linear}} function on ${\cal H}$),  
is defined, and it is shown that 
$\psi^{\cos} (f_{\mathbb R})$ is 
 a symmetric operator on ${\cal H}$.

Similarly, for each {\it{real valued}} $f_{\mathbb R} \in  {\cal S}({\mathbb R}^d \to {\mathbb R})$, (multiplicative) operators $\sqrt{-1} (\psi^{+} (f_{\mathbb R})  - \psi^{-} (f_{\mathbb R}))$, denoted by 
$\psi^{\sin} (f_{\mathbb R})$, is defined and it is shown that 
such $\psi^{\sin} (f)$ are symmetric operators on ${\cal H}$, and 
$\psi^{\sin} \, : 
{\cal S}({\mathbb R}^d \to {\mathbb C}) \ni 
f_{\mathbb C} 
\longmapsto
 \psi^{\sin} (f_{\mathbb C})$
is an (multiplicative) operator valued {\it{linear}} function on ${\cal H}$),
is 
defined, and it is shown that 
$\psi^{\sin} (f_{\mathbb R})$ is 
 a symmetric operator on ${\cal H}$.

As a consequence (see Theorem 2 and 3), for the closure (in ${\cal H}$) of 
$D^{\cos}$ the finite 
linear combinations of the multiplications of $\psi^{\cos} (f_{\mathbb R})'s$, 
denoted by ${\cal H}^{\cos}$, and  for the closure (in ${\cal H}$) of 
$D^{\sin}$ the finite 
linear combinations of the multiplications of $\psi^{\sin} (f_{\mathbb R})'s$, 
denoted by ${\cal H}^{\sin}$, 
each of the fields $<{\cal H}^{\cos}, U, \psi^{\cos}, D^{\cos}>$ and 
$<{\cal H}^{\sin}, U, \psi^{\sin}, D^{\sin}>$ 
satisfies the all of the 
G{\aa}rding-Wightman Axioms. 
More explicitly, 
in Theorem 3 and  Remarks3.3, 3.4 and 3.5, it is shown that 
in case when the underlining random field 
used 
for the present construction is 
the centered {\it{real Gaussian}} random field on ${{\mathbb R}^d}$, then the fields 
$<{\cal H}^{\cos}, U, \psi^{\cos}, D^{\cos}>$ and 
$<{\cal H}^{\sin}, U, \psi^{\sin}, D^{\sin}>$ are 
{\it{exact}} Wightman quantum fields which are 
subspaces of the original Wightman {\it{free field}} (i.e. known as the {\it{trivial quantum field}}) with $d$-dimensional space-time, 
while, in case when the underlining random field used for the present construction is 
the {\it{non Gaussian  real L{\'e}vy}} random fields on ${{\mathbb R}^d}$, then the fields 
$<{\cal H}^{\cos}, U, \psi^{\cos}, D^{\cos}>$ and 
$<{\cal H}^{\sin}, U, \psi^{\sin}, D^{\sin}>$ are {\it{non trivial exact}} Wightman quantum fields with the space time dimension $d$.

It is also shown (see Remark 3.3) that $\psi^{\cos}$ and $\psi^{\sin}$ 
 naturally correspond to a composition of {\it{creation}} and {\it{annihilation}} operators.

Another fields,  e.g., denoted by 
$<{\cal H}^{\cos, \sin}, U, \psi^{\cos, \sin}, D^{\cos, \sin}>$ and other fields 
satisfying all of the 
G{\aa}rding-Wightman Axioms,
which are super fields of 
$<{\cal H}^{\cos}, U, \psi^{\cos}, D^{\cos}>$ and $<{\cal H}^{\sin}, U, \psi^{\sin}, D^{\sin}>$ and sub fields of 
$<{\cal H}, U, \psi, D>$.

A little precisely, $\psi^+ (f)$ and  
 $\psi^- (f)$ are,  
 respectively,
defined through the {\it{Fourier}} and {\it{Fourier inverse}} transform of 
\begin{equation*}
j_{P+}^{\gamma}(\tau, \xi_1, \dots, \xi_{d-1}) = \left\{
     \begin{array}{ll}
       \displaystyle{\big(\tau^2 -( \sum_{k=1}^{d-1} \xi_k^2  + m^2) \big)^{- \gamma} }, 
 &
          \, \,  \tau > (\sum_{k=1}^{d-1} \xi_k^2 + m^2)^{\frac12}, \\
[0.2cm]
     \displaystyle{0,} & \, \, {\mbox{otherwise}},
\end{array} \right.
\end{equation*}
for 
{\textcolor{blue}{
$\gamma \in (0, \frac12)$, see Lemma 3.1 and 3.2,}}
 ($P+$ denotes the {\it{restricted Poincar{\'e} group}}), 
accompanied with the {\it{L{\'e}vy}} or the centered {\it{Gaussian}} fields $\mu$ on the {\it{real}} distribution space ${\cal S}'({\mathbb R}^d \to {\mathbb R})$, of which characteristic functions are (in the sense of {\it{Bochner-Minlos theorem}}) 
{\textcolor{blue}{
$$
\int_{{\cal S}'({\mathbb R}^d \to {\mathbb R})}
e^{i < g, \phi>} d \mu(\phi)
= e^{\int_{{\mathbb R}^d} \big( \int_{{\mathbb R} \setminus \{0\}}
(e^{i s \,  g(x)} -1 ) M(ds) \big) dx}, 
$$
}}
$$
\int_{{\cal S}'({\mathbb R}^d \to {\mathbb R})}
e^{i < g, \phi>} d \mu (\phi)
= e^{- \int_{{\mathbb R}^d} | g(x)|^2 dx}, \quad {\mbox{for real}} \, \, \,  g \in {\cal S}({\mathbb R}^d \to {\mathbb R}),
$$
respectively 
{\textcolor{blue}{(see (2.1) and (2.2) given below)}}.

The present construction of $<{\cal H}, U, \psi, D>$ (denoting $\psi = \psi^+ \, {\mbox{or}} \, \psi^-$) is relatively simple and direct, not passing through the Euclidean strategies,  e.g., the Osterwalder-Schwinger Axioms or the Nelson's Axioms. To define $\psi^+ (f)$ and  
 $\psi^- (f)$
on ${\cal H}$, general L{\'e}vy fields on ${\mathbb R}^d$ are used.

By comparing the present construction with the general flamework on indefinite metric quantum field theory presented by [St93], [LSt88], to investigate and show a clear mathematical correspondence, if any, between them is very important and interesting as the subsequent researches (see Remark 3.5).

\section{Formulations, fundamental structures, main results}
{\textcolor{blue}{ 
Let $d \in {\mathbb N}$,  including the case where $d \geq 4$,  be a given space time dimension. 
For 
{\textcolor{blue}{
$\gamma \in (0, \frac12)$}}
{\footnote{
{\textcolor{blue}{
Note that for the present case where 
$\gamma \in (0, \frac12)$ may have 
only 
a correspondence to the so called {\it{generalized free field}} 
on ${\cal S}'({\mathbb R}^d \to {\mathbb C})$, and the excluded case where 
$\gamma = \frac12$ seems to have a direct correspondence with the usual {\it{free field}} on ${\cal S}'({\mathbb R}^d \to {\mathbb C})$ (see Remark 3.5). Such exclusion is caused by a technical reasons shown by the technical Lemma 3.1 and 3.2.
}}
}}
let 
$J_{P+}^{\gamma}$ be the pseudo differential operator of which symbol is given by 
\begin{equation}
j_{P+}^{\gamma}(\tau, \xi_1, \dots, \xi_{d-1}) = \left\{
     \begin{array}{ll}
       \displaystyle{\big(\tau^2 -( \sum_{k=1}^{d-1} \xi_k^2  + m^2) \big)^{- \gamma} }, 
 &
          \, \,  \tau > (\sum_{k=1}^{d-1} \xi_k^2 + m^2)^{\frac12}, \\
[0.2cm]
     \displaystyle{0,} & \, \, {\mbox{otherwise}},
\end{array} \right. 
\end{equation}
where $\tau, \, \xi_k \, \in {\mathbb R}, \, \, k=1,\dots, d-1$. 
}}

Suppose that the underlying 
{\it{real valued}} 
 random field (a generalized white noise) $\phi$ is $\phi_{Levy}$, a Poisson type Levy  field,  or $\phi_{Gauss}$, the centered Gaussian random field, on ${\cal S}'({\mathbb R}^d \to {\mathbb R})$. 
The probability measure $\mu= \mu_{Levy}$ 
 of $\phi_{Levy}$ 
on ${\cal S}'({\mathbb R}^d \to {\mathbb R})$ is  characterized by 
a Levy measure $M(ds)\times dx$ on ${\mathbb R}\times {\mathbb R}^d$. 
{\textcolor{blue}{
Assume in particular that $M$ is a symmetric measure on ${\mathbb R}$ and 
\begin{equation}
\int_{- \infty}^{+ \infty} s^{2n +1} \, M(ds) = 0,  \qquad \int_{- \infty}^{+ \infty} s^{2n} \, M(ds) < K^{2n}, \, \, {\mbox {for some $K< \infty, \,\,  \forall n \in {\mathbb N}$}}.
\end{equation}}}
Correspondingly, 
the probability measure of  the centered Gaussian field $\phi_{Gauss}$ is denoted by $\mu= \mu_{Gauss}$
Precisely, 
by  Bochner-Minlos theorem, 
the probability space 
$({\cal S}'({\mathbb R}^d \to {\mathbb R}), {\cal B}({\cal S}'({\mathbb R}^d \to {\mathbb R})), \mu)$
 is defined through 
 {\textcolor{blue}{
\begin{equation}
\int_{{\cal S}'({\mathbb R}^d \to {\mathbb R})}
e^{i < g, \phi>} d \mu(\phi)
= e^{ \int_{{\mathbb R}^d} \big( \int_{{\mathbb R} \setminus \{0\}}
(e^{i s \,  g(x)} -1 ) M(ds) \big) dx}, \quad {\mbox{for real}} \, \, \,  g = g_{\mathbb R} \in {\cal S}({\mathbb R}^d \to {\mathbb R}),
\end{equation}
}}
for $\phi = \phi_{Levy}$ and $\mu = \mu_{Levy}$, and 
\begin{equation*}
\int_{{\cal S}'({\mathbb R}^d \to {\mathbb R})}
e^{i < g, \phi>} d \mu (\phi)
= e^{- \int_{{\mathbb R}^d} | g(x)|^2 dx}, \quad {\mbox{for real}} \, \, \,  g = g_{\mathbb R}\in {\cal S}({\mathbb R}^d \to {\mathbb R}),
\end{equation*}
for $\phi = \phi_{Gauss}$ and $\mu = \mu_{Gauss}$, respectively.


{\textcolor{blue}{
As above, 
under the assumption (2.2) for the {\it{Levy measure}} $M$, through the elementary arguments by means of the absolute convergence of the power series for the exponent 
of the right hand side of (2.3), the well-definedness of (2.3) (also, obviously for the Gaussian case) follows. 
Moreover, 
passing through such evaluations on the regularities of the characteristic functions, by 
{\it{Bochner Minlos's theorem}}, we can find (not sharp, however) the support properties of the probability measure  $\mu = \mu_{Levy}$ 
on ${\cal S}'({\mathbb R}^d \to {\mathbb R})$ 
(see [Hida 1980], and see, e.g., 
[A,Kag,Yah,YoMW2021] and [A,Kag,Kawas,Yah.,YoMW2022]).
}}

Since, the real valued random fields $\mu_{Levy}$ and $\mu_{Gauss}$ 
considered here are fundamental and simple, 
in order to get their concrete properties,  it is 
better to use the {\it{L{\'e}vy's decomposition 
theorem}}, obviously extended to random field formulation (cf. e.g. [Ito K], where the corresponding theorem of one dimensional stochastic process is introduced) (see Technical Remark 3.1).

{\textcolor{blue}{
In the sequel, every random variable is define on the probability space \\
$({\cal S}'({\mathbb R}^d \to {\mathbb R}), {\cal B}({\cal S}'({\mathbb R}^d \to {\mathbb R}), \mu)$, 
with $\mu = \mu_{Levy}$ or $\mu_{Gauss}$, where ${\cal B}({\cal S}'({\mathbb R}^d \to {\mathbb R})$ 
is the Borel $\sigma$-field of the space of real distributions ${\cal S}'({\mathbb R}^d \to {\mathbb R})$.
}}

{\textcolor{blue}{
Next, for the {\it{complex valued}} test functions $f_{\mathbb C} \in {\cal S}({\mathbb R}^d \to {\mathbb C})$, let
}} 
{\textcolor{blue}{
\begin{equation*}
(J_{P+}^{\gamma +} f_{\mathbb C})(x) 
\, 
\stackrel{\mathrm{def}}{=}
\,
 {\cal F}^{-1} (j_{P+}^{\gamma} {\hat {f_{\mathbb C}}})(x),
\end{equation*}
\begin{equation}
(J_{P+}^{\gamma -} f_{\mathbb C})(x) 
\, 
\stackrel{\mathrm{def}}{=}
\,
 {\cal F} (j_{P+}^{\gamma} {\hat {f_{\mathbb C}}})(x),
\end{equation}
}}
where ${\cal F}^{-1}$ and ${\cal F}$ are the {\it{Fourier inverse transform}} and {\it{Fourier transform}} respectively, and ${\hat{f_{\mathbb C}}}$ is the {\it{Fourier transform}} of $f_{\mathbb C}$. 

{\textcolor{blue}{
By (2.1),
for each $f_{\mathbb C} \in {\cal S}({\mathbb R}^d \to {\mathbb C})$ and $\forall \gamma \in (0, \frac12)$, 
 it holds that (more precisely see Lemma 3.1)
\begin{equation}
(J_{P+}^{\gamma +} f_{\mathbb C}), \, (J_{P+}^{\gamma -} f_{\mathbb C}) \in 
\bigcap_{2 \leq p \leq \infty}
L^p({\mathbb R}^d),
\end{equation}
(the case where $\gamma = \frac12$ is postopned to forthcoming considerations). 
Then,
for $\phi = \phi_{Levy}$ or $\phi = \phi_{Gauss}$, the ${\cal S}'({\mathbb R}^d \to {\mathbb R})$-valued random variable defined by (2.2), (2.3) and (2.4), it is possible to define the dualization 
(precisely, as a {\it{stochastic extension}} of the {\it{dualization}}, see Lemma 3.2) 
such that 
\begin{equation*}
<J_{P+}^{\gamma +} f_{\mathbb C}, \phi>, \quad f \in {\cal S}({\mathbb R}^d \to {\mathbb C}),
\end{equation*}
\begin{equation}
<J_{P+}^{\gamma -} f_{\mathbb C}, \phi>, \quad f \in {\cal S}({\mathbb R}^d \to {\mathbb C}),
\end{equation}
respectively.
}}

{\textcolor{blue}{
By the integrabilities of these random variables, it is possible to see that the  {\it{complex valued}} random variables 
$<J_{P+}^{\gamma} f_{\mathbb C}, \phi>$
satisfy and are characterized by (see Lemms 3.2 again)
}}
{\textcolor{blue}{
\begin{equation}
\int_{{\cal S}'({\mathbb R}^d \to {\mathbb R})}
e^{i < J^{\gamma}_{P+} f_{\mathbb C}, \phi>} d \mu(\phi)
= e^{ \int_{{\mathbb R}^d} \big( \int_{{\mathbb R} \setminus \{0\}}
(e^{i s (J^{\gamma}_{P+} f_{\mathbb C})(x)} -1 ) M(ds) \big) dx},
\end{equation}
}}
for $\phi = \phi_{Levy}$ and $\mu = \mu_{Levy}$, and 
\begin{equation}
\int_{{\cal S}'({\mathbb R}^d \to {\mathbb R})}
e^{i < J^{\gamma}_{P+} f_{\mathbb C}, \phi>} d \mu(\phi)
=  e^{- \int_{{\mathbb R}^d} |
 (J^{\gamma}_{P+} f_{\mathbb C})(x) |^2 dx}
\end{equation}
for $\phi = \phi_{Gauss}$ and $\mu = \mu_{Gauss}$, respectively, 
where $J^{\gamma}_{P+}$ denotes the $J^{\gamma +}_{P+}$ or $J^{\gamma -}_{P+}$.

  \bigskip
\bigskip
\begin{defi} 
Let $\phi = \phi_{Levy}$ with $\mu = \mu_{Levy}$,  or $\phi = \phi_{Gauss}$ with $\mu = \mu_{Gauss}$.
Through the formulas (2.1)-(2.7), we give the following definitions: \\
i) \quad 
 for $f_{\mathbb C} \in {\cal S} ({\mathbb R}^d \to {\mathbb C})$, 
on the probablity space 
$({\cal S}'({\mathbb R}^d \to {\mathbb R}), {\cal B}({\cal S}'({\mathbb R}^d \to {\mathbb R})), \mu)$
let 
\begin{equation}
\psi^+ (f_{\mathbb C}) \, 
 \, 
\stackrel{\mathrm{def}}{=}
\,
< J^{\gamma +}_{P+} f_{\mathbb C}, \phi> \qquad {\mbox{and}} \qquad 
\psi^- (f_{\mathbb C}) \, 
 \, 
\stackrel{\mathrm{def}}{=}
\,
< J^{\gamma -}_{P+} f_{\mathbb C}, \phi>.
\end{equation} 
{\textcolor{blue}{
Equivalently (cf. (2.6) and Lemma 3.2), as a formulation by means of the distribution valued random fields (or distribution valued random variables) $\psi^+$ and $\psi^-$ defined by (2.11) bellow, on 
$({\cal S}'({\mathbb R}^d \to {\mathbb R}), {\cal B}({\cal S}'({\mathbb R}^d \to {\mathbb R})), \mu)$}}, 
\begin{equation}
\psi^+ (f_{\mathbb C}) = <f_{\mathbb C}, \, \psi^+>, 
\qquad
\psi^- (f_{\mathbb C}) = <f_{\mathbb C}, \, \psi^->,
\end{equation}
with 
{\textcolor{blue}{
$$
\psi^+ 
\, 
\stackrel{\mathrm{def}}{=}
\,
 <\Re(J^{\gamma +}_{P +}(\cdot)), \phi> - \, i  <\Im(J^{\gamma +}_{P +}(\cdot)), \phi>, 
$$
\begin{equation}
\psi^- 
\, 
\stackrel{\mathrm{def}}{=}
\,
 <\Re(J^{\gamma +}_{P +}(\cdot)), \phi> + \, i  <\Im(J^{\gamma +}_{P +}(\cdot)), \phi>,
\end{equation}
}}
and
$$
\Re(J^{\gamma +}_{P +}(\cdot)) 
= (\sqrt{2 \pi})^{- \frac{d}2} \int_{{\mathbb R}^d} \left(
\int_{{\mathbb R}^d}
\cos((t-\cdot)\tau + ({\vec x} -\cdot) \xi)  \,
j^{\gamma +}_{P +} \, d \tau d \xi \right)
dt d{\vec x},
$$
\begin{equation}
\Im(J^{\gamma +}_{P +}(\cdot)) 
= (\sqrt{2 \pi})^{- \frac{d}2} \int_{{\mathbb R}^d} \left(
\int_{{\mathbb R}^d}
\sin((t-\cdot)\tau + ({\vec x} -\cdot) \xi) \, 
j^{\gamma +}_{P +} \,  d \tau d \xi \right)
dt d{\vec x},
\end{equation}
where $i = \sqrt{-1}$, $\xi = (\xi_1, \dots, \xi_{d-1})$ and ${\vec x} = (x_1, \dots, x_{d-1})$.\\
ii) \quad 
let ${\cal H}$ be a Hilbert space which is the completion of the 
linear combinations of 
vectors composed by the 
random variables on 
$({\cal S}', {\cal B}({\cal S}'), \mu)$ such that
\begin{equation}
 c_0 + \prod_{k= 1}^n c_k \psi (f_{{\mathbb C},k}), \quad {\mbox{for}} \quad c_0, \, c_k \in {\mathbb C},
\, \, f_{{\mathbb C},k} \in {\cal S} ({\mathbb R}^d \to {\mathbb C}),  \, \, k=1, \dots, n, \, \, n \in {\mathbb N},
\end{equation}
where $\psi (f_{{\mathbb C},k})$ denotes $\psi^+ (f_{{\mathbb C},k})$ or $\psi^- (f_{{\mathbb C},k})$, 
for which the inner product 
$(\cdot, \cdot)_{\cal H}$
is given by 
($c'_0, \, c'_l \in {\mathbb C},
 \, g_{{\mathbb C},l} \in {\cal S} ({\mathbb R}^d \to {\mathbb C}),   \, l=1, \dots, m, \, m \in {\mathbb N}$),
\begin{equation}
\big(c_0 + \prod_{k= 1}^n c_k \psi (f_{{\mathbb C},k}), c'_0 + \prod_{l= 1}^m c'_k \psi (g_{{\mathbb C},l}) \big)_{\cal H}
\stackrel{\mathrm{def}}{=}
\int_{{\cal S}' ({\mathbb R}^d \to {\mathbb R})} \big(c_0 + \prod_{k= 1}^n c_k \psi (f_{{\mathbb C},k}) \big)
\cdot 
\big({\overline{c'_0 + \prod_{l= 1}^m c'_k \psi (g_{{\mathbb C},l})}}) \big)d \mu(\phi).
\end{equation}
iii) \quad $\psi^+ (f_{\mathbb C})$ and $\psi^- (f_{\mathbb C})$, $f_{\mathbb C} \in {\cal S} ({\mathbb R}^d \to {\mathbb C})$, are the field operatos, ${\cal H}$ is the physical Hilbert space, the vacuum in ${\cal H}$ is $1 \in {\mathbb C}$, 
{\textcolor{blue}{
the invariant domain for the field operators is the linear space composed by (2.13), 
}}
and $U$ (the strongly continuous unitary representation on ${\cal H}$ of the restricted Poincar{\'e} group $(a, \Lambda) \in {\cal P}^{\uparrow}_+$ with 
$a \in {\mathbb R}^d$, $\Lambda \in {\cal L}_+^{\uparrow}$, the restricted Lorentz group on ${\mathbb R}^d$, cf., e.g. section IX.8 of [R,S], 
{\textcolor{blue}{
and see Appendix   for a detailed description on {\it{Lorentz group}} denoted by ${\cal L}$}}) is defined and characterized (for the corresponding certification see Theorem 1 below) by 
\begin{equation}
U(a, \Lambda) \psi (f_{\mathbb C} (\cdot)) 
\stackrel{\mathrm{def}}{=}
\psi (f_{\mathbb C}({\Lambda}^{-1}(\cdot -a)),
\end{equation}
where $\psi (f_{\mathbb C})$ denotes $\psi^+ (f_{\mathbb C})$ or $\psi^- (f_{\mathbb C})$, \\
iv) \quad by iii) above, an Hermitian scalar quantum field theory, in a relaxed sense (see Theorem 1-iv,  below),  
is dented by
\begin{equation}
({\cal H}, U, \psi, D),
\end{equation}
where $\psi$ denotes $\psi^+$ or $\psi^-$.
\end{defi}

\bigskip

\begin{theorem}[Theorem on fundamental structures]
Let $\phi = \phi_{Levy}$ with $\mu = \mu_{Levy}$,  or $\phi = \phi_{Gauss}$ with $\mu = \mu_{Gauss}$.
For the operator 
{\textcolor{blue}{
$U(a, \Lambda)$ defined by (2.15), }}
it holds that\\
i)
\begin{equation}
{\mbox{
$U(a, \Lambda)$ for $(a, \Lambda) \in {\cal P}^{\uparrow}_+$ is a unitary operator on ${\cal H}$, 
}}
\end{equation}
{\textcolor{blue}{
ii) \quad 
Let $\psi (f_{\mathbb C})$ denote $\psi^+ (f_{\mathbb C})$ or $\psi^- (f_{\mathbb C})$, and 
let $P_0, \, P_1, \dots, P_{d-1}$ be the self-adjoint operators on ${\cal H}$ such that 
\begin{equation}
\big(\psi (f_{\mathbb C}), U({\mathbf a}, I)\psi(f_{\mathbb C})  \big)_{\cal H} = \big(\psi(f_{\mathbb C}), \exp(i ( a_0 \cdot P_0 + \sum_{k=0}^{d-1} a_k \cdot (-P_k))) \psi(f_{\mathbb C}) 
\big)_{\cal H},
\end{equation}
}}
where ${\mathbf a} = (a_0, a_1, \dots, a_{d-1}) \in {\mathbb R}^d$, 
then the spectrums of $P_k$, $k=0, \dots, d-1$, stay in the closed forward light cone 
\begin{equation}
{\overline{V_+}} \equiv \Big\{ (\tau, \xi_1, \dots, \xi_{d-1} \, | \, \tau^2 - (\sum_{k=1}^{d-1} 
\xi_k^2)  \, \geq \, 0, \, \,  \tau \geq 0  \, \Big\},
\end{equation}
{\textcolor{blue}{
iii) \quad 
(2.18) admits the expression such that
\begin{equation}
\big(\psi(f_{\mathbb C}), \exp(i (a_0 \cdot P_0 + \sum_{k=1}^{d-1} a_k \cdot (-P_k))) \psi(f_{\mathbb C}) 
\big)_{\cal H}
= \int_{{\mathbb R}^d} e^{i (a_0 \cdot \tau + \sum_{k=1}^{d-1} a_k \cdot (-\xi_k))}
\, d\big(\psi(f_{\mathbb C}), E_{\lambda} \psi(f_{\mathbb C}) \big)_{\cal H},
\end{equation}
where $\lambda = (\tau, \xi_1, \dots, \xi_{d-1})$ and $E_{\lambda}$ is a projection-valued measure on ${\cal H}$.
}}\\
iv) \quad Consequently, 
$<{\cal H}, U, \psi, D>$ 
 satisfies all the requirements of the the G{\aa}rding-Wightman Axioms, except 
 the requirement such 
that the field operators $\psi (f)$ with real valued $f \in  {\cal S}({\mathbb R}^d \to {\mathbb R})$ are symmetric operators on the physical Hilbert space ${\cal H}$. 
\end{theorem}

{\textcolor{blue}{
By (2.10), (2.11) and (2.12),
}} define
\begin{equation}
{\psi}^{\cos} \, 
\stackrel{\mathrm{def}}{=}
\, \psi^{+} \, + \, \psi^{-}, \qquad {\psi}^{\sin} \, 
\stackrel{\mathrm{def}}{=}
\, i \, (\psi^{+} \, - \, \psi^{-}).
\end{equation}

\begin{theorem}[Main results, composed symmetric operators] 
Let ${\psi}^{\cos}$ and $\psi^{\sin}$ be the ${\cal S}'({\mathbb R}^d \to {\mathbb C})$ valued random variables defined on the probability space 
$({\cal S}'({\mathbb R}^d \to {\mathbb R}), {\cal B}({\cal S}'({\mathbb R}^d \to {\mathbb R}), \mu)$, 
{\textcolor{blue}{
with $\mu = \mu_{Levy}$ or $\mu_{Gauss}$ (see (2.21)). 
}}\\
i) \quad For real valued  function $f_{\mathbb R} \in {\cal S}({\mathbb R}^d \to {\mathbb R})$, 
{\textcolor{blue}{
by (2.10) and (2.21) it holds that 
}}
\begin{equation}
\psi^{\cos}(f_{\mathbb R}) = 2\,  \big< f_{\mathbb R}, \, <\Re(J^{\gamma +}_{P +}(\cdot)), \psi> \big>,
\end{equation}
$\psi^{\cos} (f_{\mathbb R})$ is 
 a symmetric operator on ${\cal H}$, and 
\begin{equation}
 \psi^{\cos} \, : 
{\cal S}({\mathbb R}^d \to {\mathbb C}) \ni 
f_{\mathbb C} 
\longmapsto
 \psi^{\cos} (f_{\mathbb C})
\end{equation}
is an (multiplicative) operator valued {\it{linear}} function on ${\cal S}({\mathbb R}^d \to {\mathbb C})$.\\
ii) \quad For real valued  function $f_{\mathbb R} \in {\cal S}({\mathbb R}^d \to {\mathbb R})$, 
{\textcolor{blue}{
by (2.10) and (2.21) it holds that 
}}
\begin{equation}
\psi^{\sin}(f_{\mathbb R}) = 2  \,  \big< f_{\mathbb R}, \, <\Im(J^{\gamma +}_{P +}(\cdot)), \psi> \big>,
\end{equation}
$\psi^{\sin} (f_{\mathbb R})$ is 
 a symmetric operator on ${\cal H}$, and 
\begin{equation}
 \psi^{\sin} \, : 
{\cal S}({\mathbb R}^d \to {\mathbb C}) \ni 
f_{\mathbb C} 
\longmapsto
 \psi^{\sin} (f_{\mathbb C})
\end{equation}
is an (multiplicative) operator valued {\it{linear}} function on ${\cal S}({\mathbb R}^d \to {\mathbb C})$. \\
iii) \quad Generally, for complex valued $f_{\mathbb C} \in {\cal S}({\mathbb R}^d \to {\mathbb C})$, 
{\textcolor{blue}{
by using the notation (2.10), 
}}
\begin{equation}
\psi^{\pm} (f_{\mathbb C}) 
\, 
\stackrel{\mathrm{def}}{=}
\,
\psi^{+}(f_{\mathbb C}) + \psi^{-}({\bar{f}}_{\mathbb C}), \quad 
{\mbox{and}}
\quad
\psi^{\mp} (f_{\mathbb C}) 
\, 
\stackrel{\mathrm{def}}{=}
\,
i \, (\psi^{+}(f_{\mathbb C}) - \psi^{-}({\bar{f}}_{\mathbb C})),
\end{equation}
are, 
 on ${\cal H}$, the 
(multiplicative) symmetric operator valued linear function on ${\cal S}({\mathbb R}^d \to {\mathbb C})$.
\end{theorem}

Next, 
let the linear spaces $D^{\cos}$, $D^{\sin}$, $D^{\cos, \sin}$, $D^{\pm}$, $D^{\mp}$ and $D^{\pm,\mp}$ be such that $$
 \quad {\mbox{for}} \quad c_0, \, c_k \in {\mathbb C},
\, \, f_{{\mathbb C},k} \in {\cal S} ({\mathbb R}^d \to {\mathbb C}),  \, \, k=1, \dots, n, \, \, n \in {\mathbb N},$$
\begin{equation} D^{\cos} 
\, 
\stackrel{\mathrm{def}}{=}
\,
{\mbox{The finite linear combinations of the vectors} }\, \, 
 c_0 + \prod_{k= 1}^n c_k \psi^{\cos} (f_{{\mathbb C},k});
\end{equation}
\begin{equation} D^{\sin} 
\, 
\stackrel{\mathrm{def}}{=}
\,
{\mbox{The finite linear combinations of the vectors} }\, \, 
 c_0 + \prod_{k= 1}^n c_k \psi^{\sin} (f_{{\mathbb C},k});
\end{equation}
\begin{equation} D^{\cos, \sin} 
\, 
\stackrel{\mathrm{def}}{=}
\,
{\mbox{The finite linear combinations of the vectors} }\, \, 
 c_0 + \prod_{k= 1}^n c_k \psi (f_{{\mathbb C},k}),
\end{equation}
where $\psi = \psi^{\cos} \, \mbox{or} \, \, \psi^{\sin}$; 
\begin{equation} D^{\pm} 
\, 
\stackrel{\mathrm{def}}{=}
\,
{\mbox{The finite linear combinations of the vectors} }\, \, 
 c_0 + \prod_{k= 1}^n c_k \psi^{\pm} (f_{{\mathbb C},k});
\end{equation}
\begin{equation} D^{\mp} 
\, 
\stackrel{\mathrm{def}}{=}
\,
{\mbox{The finite linear combinations of the vectors} }\, \, 
 c_0 + \prod_{k= 1}^n c_k \psi^{\mp} (f_{{\mathbb C},k});
\end{equation}
\begin{equation} D^{\pm, mp} 
\, 
\stackrel{\mathrm{def}}{=}
\,
{\mbox{The finite linear combinations of the vectors} }\, \, 
 c_0 + \prod_{k= 1}^n c_k \psi (f_{{\mathbb C},k}),
\end{equation}
where $\psi = \psi^{\pm} \, \mbox{or} \, \, \psi^{\mp}$.

\begin{theorem}[Main results, the constructed exact Wightman fields]
Let ${\cal H}^{\cos}$, ${\cal H}^{\sin}$, ${\cal H}^{\cos, \sin}$, ${\cal H}^{\pm}$, ${\cal H}^{\mp}$ 
and ${\cal H}^{\pm, \mp}$ be respectively the Hilbert spaces which are completion of 
${D}^{\cos}$, $D^{\sin}$, ${D}^{\cos, \sin}$, ${D}^{\pm}$, ${D}^{\mp}$ 
and ${D}^{\pm, \mp}$ respectively,
 with respect to the norm corresponding to the 
 {\textcolor{blue}{
 inner product given by (2.14).  
 }}
Then, 
each of the fields
\begin{equation}
 <{\cal H}^{\cos}, U, \psi^{\cos}, D^{\cos}>, \quad  
<{\cal H}^{\sin}, U, \psi^{\sin}, D^{\sin}>, \quad 
<{\cal H}^{\cos, \sin}, U, \psi^{\cos, \sin}, D^{\cos, \sin}>
\end{equation}
\begin{equation}
<{\cal H}^{\pm}, U, \psi^{\pm}, D^{\pm}>, \quad 
<{\cal H}^{\mp}, U, \psi^{\mp}, D^{\mp}>, \quad 
<{\cal H}^{\pm, \mp}, U, \psi^{\pm, \mp}, D^{\pm, \mp}>,
\end{equation}
satisfies the all of the 
G{\aa}rding-Wightman Axioms.

Moreover,  it holds that
\begin{equation}
{\cal H}^{\cos}, \, {\cal H}^{\sin} \subset {\cal H}^{\cos, \sin} \subset {\cal H}^{\pm, \mp},
\end{equation}
\begin{equation}
{\cal H}^{\pm}, \, {\cal H}^{\mp}  \subset {\cal H}^{\pm, \mp} \subset  {\cal H}.
\end{equation}
\end{theorem}

{\bf{Sketch of the proof for Theorem 1}} \\
Let $J^{\gamma}_{P+}$ denote the $J^{\gamma +}_{P+}$ or $J^{\gamma -}_{P+}$.
Here, we give a sketch of the proof for 
{\textcolor{blue}{
(2.17), the Poincare invariance property of the fields. 
}}
Since the Fourier transform of the fields, the corresponding characteristic functions (in the present case they are {\it{Fourier Laplace transforms}} for complex valued $J^{\gamma}_{P+} f_{\mathbb C}$,  
{\textcolor{blue}{
(2.8) and (2.9) characterize the moment properties of $\psi(f_{\mathbb C})$ completely, 
and the {\it{inner product}} of the Hilbert space ${\cal H}$ is defined through (2.14), 
 to show that 
(2.17) holds, it suffices to see that (below, for a notational simplicity we prove (2.17) for $U(-a, {\Lambda}^{-1})$)
}}
\begin{eqnarray}
\lefteqn{
\int_{{\mathbb R}^d} \big( \int_{{\mathbb R} \setminus \{0\}}
(e^{i s (J^{\gamma}_{P+} f_{\mathbb C}(a + \Lambda \cdot)
+ {\overline{J^{\gamma}_{P+} g_{\mathbb C} (a + \Lambda \cdot)}}
)(x)} -1 ) M(ds) \big) dx
} \nonumber \\
&&=
\int_{{\mathbb R}^d} \big( \int_{{\mathbb R} \setminus \{0\}}
(e^{i s (J^{\gamma}_{P+} f_{\mathbb C}(\cdot)
+ {\overline{J^{\gamma}_{P+} g_{\mathbb C} (\cdot)}}
)(x)} -1 ) M(ds) \big) dx,
\end{eqnarray}
for any $(a, \Lambda) \in {\cal P}^{\uparrow}_+$, $f_{\mathbb C},\,  g_{\mathbb C} \in {\cal S} ({\mathbb R}^d \to {\mathbb C})$, for $\phi_{Levy}$ and $\mu_{Levy}$, and 
\begin{equation}
\int_{{\mathbb R}^d} |
 (J^{\gamma}_{P+} f_{\mathbb C}(a + \Lambda \cdot)
+ {\overline{J^{\gamma}_{P+} g_{\mathbb C} (a + \Lambda \cdot)}}
)(x) |^2 dx
=
\int_{{\mathbb R}^d} |
 (J^{\gamma}_{P+} f_{\mathbb C}(\cdot)
+ {\overline{J^{\gamma}_{P+} g_{\mathbb C} ( \cdot)}}
)(x) |^2 dx, 
\end{equation}
for any $(a, \Lambda) \in {\cal P}^{\uparrow}_+$, $f_{\mathbb C},\,  g_{\mathbb C} \in {\cal S} ({\mathbb R}^d \to {\mathbb C})$, for $\phi_{Gauss}$ and $\mu_{Gauss}$ respectively.

{\textcolor{blue}{
{\it{For the present formulation, (3.37) holds, generally,  for any additive field (process), hence for any Levy random field on ${\mathbb R}^d$ and for the matrix $\Lambda$ such that 
$|\Lambda| =1$ (e.g., for $\Lambda \in {\cal L}_+^{\uparrow}$, see Appendix).}}
In fact,  for $a=0$, (2.38) can be shown as follows (then, the case where $a \ne 0$ will be clear):
}} \\
{\textcolor{blue}{
for $\Lambda \in {\cal L}_+^{\uparrow}$ \,
(here, we denote $\hat{f}$ as the {\it{Fourier transform}} of $f$)
}}
\begin{eqnarray}
 \lefteqn{\int_{{\mathbb R}^d} |(J^{\gamma}_{P+} f_{\mathbb C} ( \Lambda \cdot)
+ {\overline{J^{\gamma}_{P+} g_{\mathbb C} ( \Lambda \cdot)}}
)(x)|^{2} \, dx }
 \nonumber \\
&& =
\int_{{\mathbb R}^d}  \left|
\int_{{\mathbb R}^d}
\left(J^{\gamma}_{P+}(x-y) f_{\mathbb C} ( \Lambda y)
+ {\overline{J^{\gamma}_{P+}(x-y) g_{\mathbb C} ( \Lambda y)}}
\right)
dy
\right|^{2} \, dx
\nonumber \\
 && = \int_{{\mathbb R}^d} \Big| \int_{{\mathbb  R}^d} \Big( \int_{{\mathbb R}^d}
 j_{P+}^{\gamma}(\xi) e^{i(x-y) \cdot \xi} ( \int_{{\mathbb R}^d} e^{i \eta \cdot (\Lambda y)}
 {\hat{f_{\mathbb C}}}(\eta) d \eta ) dy \Big) d \xi 
 \nonumber \\
&& + \int_{{\mathbb  R}^d} \Big( \int_{{\mathbb R}^d}
 j_{P+}^{\gamma}(\xi) e^{-i(x-y) \cdot \xi} ( \int_{{\mathbb R}^d} e^{-i \eta \cdot (\Lambda y)}
{\overline{ {\hat{g_{\mathbb C}}}(\eta)}} d \eta ) dy \Big) d \xi \Big|^{2} dx 
\nonumber \\
 && =
 \int_{{\mathbb R}^d} \Big| \int_{{\mathbb  R}^d} 
 j_{P+}^{\gamma}(\xi) e^{i x \cdot \xi} 
 \Big( \int_{{\mathbb R}^d}
  ( \int_{{\mathbb R}^d} e^{i y \cdot(- \xi + {}^t \Lambda \eta)}
 {\hat{f_{\mathbb C}}}(\eta) d \eta ) dy \Big) d \xi 
 \nonumber \\
&&+
 \int_{{\mathbb  R}^d} 
 j_{P+}^{\gamma}(\xi) e^{-i x \cdot \xi} 
 \Big( \int_{{\mathbb R}^d}
  ( \int_{{\mathbb R}^d} e^{-i y \cdot(- \xi + {}^t \Lambda \eta)}
 \overline{\hat{g_{\mathbb C}}}(\eta) d \eta ) dy \Big) d \xi \Big|^{2} dx
 \nonumber \\
 &&= 
 \int_{{\mathbb R}^d} \Big| 
 \int_{{\mathbb R}^d} j_{P+}^{\gamma} ( {}^t \Lambda \eta) 
 e^{i x \cdot ({}^t \Lambda \eta)} {\hat{f_{\mathbb C}}}(\eta) d \eta 
+ \int_{{\mathbb R}^d} j_{P+}^{\gamma} ( {}^t \Lambda \eta) 
 e^{-i x \cdot ({}^t \Lambda \eta)} \overline{\hat{g_{\mathbb C}}}(\eta) d \eta \Big|^{2} dx
 \nonumber \\
 &&= 
 \int_{{\mathbb R}^d} \Big| 
 \int_{{\mathbb R}^d} j_{P+}^{\gamma} (\eta) 
 e^{i \eta \cdot (\Lambda x)} {\hat{f_{\mathbb C}}}(\eta) d \eta 
+ \int_{{\mathbb R}^d} j_{P+}^{\gamma} (\eta) 
 e^{- i \eta \cdot (\Lambda x)} \overline{\hat{g_{\mathbb C}}}(\eta) d \eta \Big|^{2} dx
  \\
 &&= 
 \int_{{\mathbb R}^d} 
 | {\Lambda}^{-1}| \, 
 \left| 
 \int_{{\mathbb R}^d} j_{P+}^{\gamma} (\eta) 
 e^{i \eta \cdot z} {\hat{f_{\mathbb C}}}(\eta) d \eta 
+
\int_{{\mathbb R}^d} j_{P+}^{\gamma} (\eta) 
 e^{-i \eta \cdot z} \overline{\hat{g_{\mathbb C}}}(\eta) d \eta
\right|^{2} dz 
 \\
 && = \int_{{\mathbb R}^d} \left| (J_{P+}^{\gamma} f_{\mathbb C})(z) 
+
(\overline{J_{P+}^{\gamma} g_{\mathbb C}})(z)
\right|^{2} dz,
 \end{eqnarray}
 {\textcolor{blue}{
where (2.39) follows from the fact that $j_{P+}^{\gamma}$ given by (2.1)  is invariant with respect to the {\it{Lorentz transform}} and 
${}^t \Lambda \in {\cal L}_+^{\uparrow}$
for 
$\Lambda \in {\cal L}_+^{\uparrow}$ (see (4.11) and the descriptions given in Appendix),
 for (2.40) we set $\Lambda x = z$, and (2.41) 
follows from the fact that $|{\Lambda}| = 1$ for $\Lambda \in {\cal L}_+^{\uparrow}$ 
(cf. the definition of the restricted Lorentz group (see Appendix)). 
By making use of 
(2.38), 
from  (2.7) we see that 
(2.17) 
holds 
 for $(a, \Lambda) \in {\cal P}^{\uparrow}_+$ with $a =0$.
Now, for $(a, \Lambda) \in {\cal P}^{\uparrow}_+$ with $a \in {\mathbb R}^d$, 
(2.17) is clear.}}

{\textcolor{blue}{
To prove (2.37), in the proof of (2.38) as above, we can replace 
$$|(J^{\gamma}_{P+} f_{\mathbb C} ( \Lambda \cdot)
+ {\overline{J^{\gamma}_{P+} g_{\mathbb C} ( \Lambda \cdot)}}
)(x)|^{2}
$$
by
$$
\big( \int_{{\mathbb R} \setminus \{0\}}
(e^{i s (J^{\gamma}_{P+} f_{\mathbb C}(\cdot)
+ {\overline{J^{\gamma}_{P+} g_{\mathbb C} (\cdot)}}
)(x)} -1 ) M(ds) \big),
$$
then, (2.37) also follows clearly.}}

{\textcolor{blue}{
We have to note, repeatedly,  that for (2.37) and (2.38), the Poincare invariance correspondences, the property of 
{\it{
stationary independent increment (with respect to the Lebesgue measure)}} of the underlining random fields (i.e., the additive fields) 
$\phi = \phi_{Levy}$ with $\mu = \mu_{Levy}$,  or $\phi = \phi_{Gauss}$ with $\mu = \mu_{Gauss}$ is crucial as the mathematical structure.
}}

 {\textcolor{blue}{
In order to see that ii) and iii), the property of spectrum of projection valued measure
((2.18), (2.19) and (2.20)), hold, we can perform a procedure by means of the {\it{Bochner theorem}} (see. for e.g., Theorem IX.9 of [Rsi1975], or Theorem 2 in Section XI-13 of [YosidaK{\^o}saku1980], or  Theorem 12.2 of [Ito S.1978] (p. 405) together with Theorem 12.7 and {\it{Stone theorem}}, Theorem 12.8 of [Ito S.1978] (p.410 and 413). In short, and roughly, firstly the left hand side of (2.18) admits the expression of its  right hand side can be  shown, and then for the right hand side of (2.18) by taking the Fourier inverse transform we have the projection valued measure whose corresponding spectrum $\lambda$ stays in ${\overline{V_+}}$ given by (2.19)
}}
 {\textcolor{blue}{(see Lemma 3.3 in detail).}}

\bsquare

\bigskip

{\bf{Sketch of the proof for Theorem 2, 3}} \\
Since, $\psi^{\cos}$, $\psi^{\sin}$, $\psi^{\cos, \sin}$, $\psi^{\pm}$, $\psi^{\mp}$ and $\psi^{\pm, \mp}$ are real valued multiplicative operators which are linear combinations of $\psi$, from Theorem 1, the results provided in Theorems 2 and 3 are clear.
\bsquare

\section{Remarks and Lemmas}
\begin{rem}[Technical Remark 1]
On the statement derived by Bochner-Minlos theorem.  
The Bochner-Minlos theorem
 is surely strong in the sense that it 
admits to be applied to general random fields on nuclear spaces, on the other hand,  the support property of the corresponding probability measures (random fields) available by the theorem is not sharp (see, e.g., [BFS], [AKYY]).
\end{rem}





\begin{rem}
Note that 
\begin{equation}
\int_{{\mathbb R}^d} ({\cal F}^{-1} j_{P+})(x-y) \cdot ({\cal F}^{-1} j_{P+})(y-z)dy
= {\rm{constant}} \cdot ({\cal F}^{-1}  (j_{P+})^2)(x-z),
\end{equation}
while, by the support property of $j_{P+}$ with respect to $\tau$ (see (1.1), we see that 
\begin{equation}
\int_{{\mathbb R}^d} ({\cal F}^{-1} j_{P+})(x-y) \cdot ({\cal F}^{-1} j_{P+})(z-y)dy
= 0. \qquad \bsquare
\end{equation}
\end{rem}

{\textcolor{blue}{
\begin{rem}
For the case such that $\phi = \phi_{Gauss}$ with $\mu = \mu_{Gauss}$, the field 
$({\cal H}, U, \psi, D)$ defined by (2.16) would be expected to be the usual {\it{free field}}, 
the original Wightman free field, 
but it is not true. It can be seen as follows: \\
By taking $\lambda = \frac12$, let 
$W$ be the Wightman function of the {\it{free field}}, such that 
for $x = (t,{\vec{x}}), \, y=(s, {\vec{y}}) \in {\mathbb R}^d$,
\begin{equation}
W(x-y) = \frac{1}{(2 \pi)^{d-1}} \int_{{\mathbb R}^{d-1}}  \left(  
e^{i(\vec{x} - \vec{y}) \cdot \xi}
\frac{\exp \Big({i(t-s){\sqrt{\sum_{k=1}^{d-1} \xi_k^2}}} \Big) }
{{\sqrt{\sum_{k=1}^{d-1} \xi_k^2}}} \right) d\xi
\end{equation}
with $\xi = \xi_1, \dots, \xi_{d-1}$.
Then, for the Wightman free field $\phi_{free}$, it holds that, for $f, \, g   
\in {\cal S} ({\mathbb R}^d \to {\mathbb C})$,
\begin{eqnarray}
\lefteqn{
(\phi_{free}(f), \phi_{free}(g))_{Wightman} = (\psi^+(f), \psi^+(g))_{\cal H} 
} \nonumber \\
&&= E^{\mu_{Gauss}}[ \psi^+(f), \cdot \overline{\psi^+(g))}]
= \int_{{\mathbb R}^{d-1}} \int_{{\mathbb R}^{d-1}} f(x) W(x-y) \overline{g}(y) dx dy,
\end{eqnarray}
where $(\, \cdot, \cdot \, )_{Wightma}$ denotes the inner product of the (original) Wightman 
free field, and  $E^{\mu_{Gauss}}[ \cdot ]$ denotes the expectation with respect to the probability measure $\mu_{Gauss}$,
while for  $f_1, \, f_2, \, g_1, \, g_2   
\in {\cal S} ({\mathbb R}^d \to {\mathbb C})$,
\begin{eqnarray}
\lefteqn{
(\phi_{free}(f_1) \phi_{free}(f_2), \phi_{free}(g_1) \phi_{free}(g_2))_{Wightman}
}\\
&& = (\int_{{\mathbb R}^{d-1}} \int_{{\mathbb R}^{d-1}} f_1(x) W(x-y) \overline{g_1}(y) dx dy) \cdot 
(\int_{{\mathbb R}^{d-1}} \int_{{\mathbb R}^{d-1}} f_2(x) W(x-y) \overline{g_2}(y) dx dy),
\nonumber \\
&& +
(\int_{{\mathbb R}^{d-1}} \int_{{\mathbb R}^{d-1}} f_1(x) W(x-y) \overline{g_2}(y) dx dy) \cdot 
(\int_{{\mathbb R}^{d-1}} \int_{{\mathbb R}^{d-1}} f_2(x) W(x-y) \overline{g_1}(y) dx dy),
\nonumber \\
&&+
(\int_{{\mathbb R}^{d-1}} \int_{{\mathbb R}^{d-1}} f_2(x) W(x-y) f_1(y) dx dy) \cdot 
(\int_{{\mathbb R}^{d-1}} \int_{{\mathbb R}^{d-1}} \overline{g_2}(x) W(x-y) \overline{g_1}(y) dx dy),
\nonumber 
\end{eqnarray}
but 
\begin{eqnarray}
\lefteqn{
(\psi^+(f_1) \psi^+(f_2), \psi^+(g_1) \psi^+(g_2))_{\cal H}
}\\
&& = (\int_{{\mathbb R}^{d-1}} \int_{{\mathbb R}^{d-1}} f_1(x) W(x-y) \overline{g_1}(y) dx dy) \cdot 
(\int_{{\mathbb R}^{d-1}} \int_{{\mathbb R}^{d-1}} f_2(x) W(x-y) \overline{g_2}(y) dx dy),
\nonumber \\
&& +
(\int_{{\mathbb R}^{d-1}} \int_{{\mathbb R}^{d-1}} f_1(x) W(x-y) \overline{g_2}(y) dx dy) \cdot 
(\int_{{\mathbb R}^{d-1}} \int_{{\mathbb R}^{d-1}} f_2(x) W(x-y) \overline{g_1}(y) dx dy).
\nonumber 
\end{eqnarray}
Since the last term of (3.6) corresponding to the one of (3.5) is 
\begin{equation}
(\int_{{\mathbb R}^{d-1}} \int_{{\mathbb R}^{d-1}} f_2(x) W(x+y) f_1(y) dx dy) \cdot 
(\int_{{\mathbb R}^{d-1}} \int_{{\mathbb R}^{d-1}} \overline{g_2}(x) W(x+y) \overline{g_1}(y) dx dy),
\end{equation}
which vanishes because of the support property of $j~{\gamma}_{P +}$ defined by (2.1) (see Remark 3.4)\\
\bsquare
\end{rem}
}}

\begin{rem}
By Remarks 3.2 and 3.3, 
on $({\cal H}, U, \psi, D)$, 
in short, 
the vectors such that 
$\psi^+ (f_{{\mathbb C}, 1}) \cdots \psi^+ (f_{{\mathbb C}, k})$, $k \in {\mathbb N}$, 
are identified as the Wick monomials, $$:\, \psi^+ (f_{{\mathbb C}, 1}) \cdots \psi^+ (f_{{\mathbb C}, k}) \, :,$$
for $\phi = \phi_{Levy}$ with $\mu = \mu_{Levy}$,  or $\phi = \phi_{Gauss}$ with $\mu = \mu_{Gauss}$. \\
\bsquare
\end{rem}

\begin{rem} 
By Remark 3.3, we see that 
for the case such that $\phi = \phi_{Gauss}$ with $\mu = \mu_{Gauss}$, the field 
$({\cal H}, U, \psi, D)$ defined by (1.12) can not be the original Wightman {\it{free field}}. Nevertheless, by Remark 3.6, the field 
$<{\cal H}^{\pm, \mp}, U, \psi^{\pm, \mp}, D^{\pm, \mp}>$ might be identified with the original  
Wightman free field.  This problem of the identification will be a subject of the subsequent papers.
\bsquare
\end{rem}

\begin{rem}
Correspondences with the theory of indefinite metric quantum fields ([AHW 1997], [J,S], [St]).
\end{rem}

{\textcolor{blue}{
\begin{lmm}[Technical Lemma 1; 
On the integrability of $J^{\gamma +}_{P+} f$ and $J^{\gamma -}_{P+} f$]
Suppose that $\gamma \in (0, \frac12)$, and $j^{\gamma}_+$ is defined by (2.1) with a given {\it{mass constant}} $m>0$, then for each $f \in {\cal S}({\mathbb R}^d \to {\mathbb C})$, there exists a constant $C \stackrel{\mathrm{def}}{=} C(d,m,f) < \infty$, depending only on $d, m, f$,  and the following holds (see (2.4) for the definition of 
$j^{\gamma}_{P+} {\hat{f}}_{\mathbb C}$, 
$J^{\gamma +}_{P+} f_{\mathbb C}$ and 
$J^{\gamma -}_{P+} f_{\mathbb C}$)
: 
\begin{equation}
\| j^{\gamma}_{P+} {\hat{f}}_{\mathbb C} \|_{L^q({\mathbb R}^d)} \leq C, \qquad \forall q \in [1, 2],
\end{equation}
\begin{equation}
\| J^{\gamma +}_{P+} f_{\mathbb C} \|_{L^p({\mathbb R}^d)} \leq C, 
\qquad \|J^{\gamma -}_{P+} f_{\mathbb C} \|_{L^p({\mathbb R}^d)} \leq C,  \qquad \forall p \in [2, \infty].
\end{equation}
\end{lmm}
}}

{\textcolor{blue}{
{\bf{Proof of Lemma 3.1}}\\
(3.9) is a direct consequence of (3.8) with {\it{Hausdorff-Young inequality}}, and for the proof of this Lemma, it suffices to show (3.8).  For each given $f \in {\cal S}({\mathbb R}^d \to {\mathbb C})$, we have to see that a universal constant $C$ by which (3.8) holds. First of all, we note that 
for $f_{\mathbb C} \in {\cal S}({\mathbb R}^d \to {\mathbb C})$ the Fourier transform $\hat{f}$ is also an element of ${\cal S}({\mathbb R}^d \to {\mathbb C})$ and for a unit square, $m$ a given mass,  
$S_m = \{(\tau, r) \, | \, -m-1 \leq \tau \leq m+1, \, 0 \leq r \leq 1 \}$ with 
$r = |\xi|$, $\xi = (\xi_1, \dots, \xi_{d-1}) \in {\mathbb R}^{d-1}$ there exists a constant $C_1< \infty$ and $C_2 < \infty$ such that
\begin{equation}
\sup_{(\tau, r) \in S_m} |\hat{f} | \leq C_1, 
\end{equation}
\begin{equation}
|\hat{f} (\tau, r)| \leq \max \{ \frac{C_2}{|\tau|^2}, \, \frac{C_2}{r^{d+1}} \}, \qquad \forall (\tau, \xi) 
\notin S_m.
\end{equation}
Here, we show an estimate corresponding to the bound (3.8) only for a region $S_m$, since the estimate on the outside of $S_m$ is easier by (3.11).
Let $\gamma \in (0, \frac12)$ and $q \in [1,2]$, then by (3.10), we see that 
\begin{eqnarray}
\lefteqn{\int_{S_m} | j^{\gamma}_{P+} (\tau, r) \, \hat{f}(\tau, r)|^q \,  r^{d-1}  d \tau d r}
\nonumber \\
&& = \int_{-1}^1 ( \int_{\sqrt{r^2 + m^2}}^{m+1} \big(\tau^2 - (r^2 + m^2) \big)^{-\gamma q} 
|\hat{f} (\tau, r) |^q d \tau ) \, r^{d-1} dr \nonumber \\
&& \leq (C_1)^q \int_{-1}^1 \Big( \int_{\sqrt{r^2 + m^2}}^{m+1}( \tau - \sqrt{r^2  + m^2})^{-\gamma q} ( \tau + \sqrt{r^2  + m^2})^{-\gamma q} d \tau \Big) \, r^{d-1} dr \nonumber \\
&& \leq C_1 \int_{-1}^1 (2 \sqrt{r^2 + m^2})^{- \gamma q} \, \big( \int_0^{m + 1 - \sqrt{r^2 + m^2}}
\tau^{-\gamma q} d \tau \big) dr \nonumber \\
&&  \leq (C_1)^q (2 m)^{- \gamma q} \frac{1}{1- \gamma q}.
\end{eqnarray}
(3.12) shows the bound (3.8) on the region $S_m$.\\
\bsquare
}}

{\textcolor{blue}{
\begin{lmm}[Technical Lemma 2; 
Well-definedness of $<J_{P+}^{\gamma +} f_{\mathbb C}, \phi>$,  
$<J_{P+}^{\gamma -} f_{\mathbb C}, \phi>$] 
On the probability space $({\cal S}'({\mathbb R}^d \to {\mathbb R}), {\cal B}({\cal S}'({\mathbb R}^d \to {\mathbb R}), \mu)$ with 
 $\mu = \mu_{Levy}$, defined by  (2.2) for a Levy measure $M$ and 
 (2.3), through {\it{Bochner-Minlos's theorem}}, 
 and through an argument by means of a stochastic extension of the dualizations
 $<g, \phi>$, $g \in {\cal S}({\mathbb R}^d \to {\mathbb C})$, 
for any  $\gamma \in (0, \frac12)$ and 
for any $f_{\cal C} \in {\cal S}({\mathbb R}^d \to {\mathbb C})$,
$<J^{\gamma}_{P+} f_{\mathbb C}, \phi>$, 
 denoting $J^{\gamma +}_{P+}$ or $J^{\gamma -}_{P+}$
  simply 
  by  $J^{\gamma}_{P+}$, 
 can be defined as a complex valued random variable satisfying 
 \begin{equation}
 <J^{\gamma}_{P+} f_{\mathbb C}, \phi> \in \bigcap_{p=2}^{\infty} L^p_{\mu},
 \end{equation}
\begin{equation}
\int_{{\cal S}'} \big(\Re(<J^{\gamma}_{P+} f_{\mathbb C}, \phi>) \big)^{2n} d\mu( \phi)
= (-1)^{n} \frac{d^{2n}}{d \lambda^{2n}} \Phi_f(\lambda)|_{\lambda = 0},
\end{equation}
\begin{equation}
\int_{{\cal S}'} \big(\Im(<J^{\gamma}_{P+} f_{\mathbb C}, \phi>) \big)^{2n} d\mu( \phi)
= (-1)^{n} \frac{d^{2n}}{d \lambda^{2n}} \Psi_f(\lambda)|_{\lambda = 0},
\end{equation}
\begin{equation}
\int_{{\cal S}'} \big(\Re(<J^{\gamma}_{P+} f_{\mathbb C}, \phi>) \big)^{2n-1} d\mu(\phi)
= \int_{{\cal S}'} \big(\Im(<J^{\gamma}_{P+} f_{\mathbb C}, \phi>) \big)^{2n-1} \mu(d \phi)
= 0, \quad n \in {\mathbb N},
\end{equation}
where 
\begin{eqnarray}
\lefteqn{
\Phi_f(\lambda) }
\nonumber \\
&&= \exp \left[ {\int_{{\mathbb R}^d} \Big[ \int_{{\mathbb R} \setminus \{0\}} 
\Big(
\cos \big({\lambda s (\Re J^{\gamma}_{P+} f_{\mathbb C})(x)} \big) 
- 1 \Big) 
\, M(ds) \Big] dx} \right], 
\nonumber \\
&&= 
\sum_{n=0}^{\infty} \frac{1}{n!} 
\left( \sum_{k=1}^{\infty} \frac{(-1)^k}{(2k)!} \Big( \int_{{\mathbb R}^d}  
\big(\lambda  (\Re J^{\gamma}_{P+} f_{\mathbb C}(x)) \big)^{2k} dx \Big) \Big(\int_{{\mathbb R} \setminus \{0\}}
 s^{2k} M(ds) \Big)
\right)^n,
\end{eqnarray}
\begin{eqnarray}
\lefteqn{
\Psi_f(\lambda)}
\nonumber \\
 &&= \exp \left[ { \int_{{\mathbb R}^d} \Big[ \int_{{\mathbb R} \setminus \{0\}} 
\Big(
\cos \big({\lambda s (\Im J^{\gamma}_{P+} f_{\mathbb C})(x)} \big) 
-1 \Big) \, M(ds) \Big] dx} \right] \nonumber \\
&&=
\sum_{n=0}^{\infty} \frac{1}{n!} 
\left(\sum_{k=1}^{\infty} \frac{(-1)^k}{(2k)!} \Big( \int_{{\mathbb R}^d}  
\big(\lambda  (\Im J^{\gamma}_{P+} f_{\mathbb C}(x)) \big)^{2k} dx \Big) \Big(\int_{{\mathbb R} \setminus \{0\}}
 s^{2k} M(ds) \Big)
\right)^n,
\end{eqnarray}
 both of which right hand side converges absolutely.
Also,  
the equation (2.7) holds explicitly such that 
\begin{eqnarray}
\lefteqn{
\int_{{\cal S}'({\mathbb R}^d \to {\mathbb R})}
e^{i < J^{\gamma}_{P+} f_{\mathbb C}, \phi>} d \mu(\phi) }  \nonumber \\
&&= \exp \left( \sum_{k=1}^{\infty} \frac{(-1)^k}{(2k)!} \Big( \int_{{\mathbb R}^d}  
\big(J^{\gamma}_{P+} f_{\mathbb C}(x) \big)^{2k} dx \Big) \Big(\int_{{\mathbb R} \setminus \{0\}}
 s^{2k} M(ds) \Big)
\right) \nonumber \\
&& =
\sum_{n=0}^{\infty} \frac{1}{n!} 
\left(\sum_{k=1}^{\infty} \frac{(-1)^k}{(2k)!} \Big( \int_{{\mathbb R}^d}  
\big(J^{\gamma}_{P+} f_{\mathbb C}(x) \big)^{2k} dx \Big) \Big(\int_{{\mathbb R} \setminus \{0\}}
 s^{2k} M(ds) \Big)
\right)^n,
\end{eqnarray}
where the righthand side converges absolutely.
\end{lmm}
}}

{\textcolor{blue}{
\begin{rem}[Truncated Wightman functions]
By, (3.14), (3.15), (3.16), (3.17) and (3.18) 
for the present quantum field model we have the truncated Wightman functions $\{W_n\}_{n \in \mathbb N}$
 (see [AGW]), which are defined through the moment functionals 
$ \displaystyle \int_{{\cal S}'} \big( <J^{\gamma}_{P+} f_{\mathbb C}, \phi>) \big)^n d\mu( \phi)$ for $f_{\cal C} \in {\cal S}({\mathbb R}^d \to {\mathbb C})$.
For simplicity, e.g., denote 
$$F(x) = 
 (\Re J^{\gamma}_{P+} f_{\mathbb C})(x) \quad 
 {\mbox{and}} \quad 
 m_{2k} = \int_{{\mathbb R} \setminus \{0\}}
 s^{2k} M(ds).
 $$
then, 
we see that 
\begin{equation*}
\int_{{\cal S}'} \big(\Re(<J^{\gamma}_{P+} f_{\mathbb C}, \phi>) \big)^{2} d\mu( \phi)
= (\int_{{\mathbb R}^d} F(x)^2 dx) m_2,
\end{equation*}
\begin{equation*}
\int_{{\cal S}'} \big(\Re(<J^{\gamma}_{P+} f_{\mathbb C}, \phi>) \big)^{4} d\mu( \phi)
= (\int_{{\mathbb R}^d} F(x)^4 dx) m_4 
+ 3 \big( (\int_{{\mathbb R}^d} F(x)^2 dx) m_2 \big)^2,
\end{equation*}
\begin{eqnarray*}
\lefteqn{
\int_{{\cal S}'} \big(\Re(<J^{\gamma}_{P+} f_{\mathbb C}, \phi>) \big)^{6} d\mu( \phi)
} \\
&&= (\int_{{\mathbb R}^d} F(x)^6 dx) m_6 
+ 15 \big((\int_{{\mathbb R}^d} F(x)^4 dx) m_4 \big)\big( (\int_{{\mathbb R}^d} F(x)^2 dx) m_2 \big)
+ 15 
\big( (\int_{{\mathbb R}^d} F(x)^2 dx) m_2 \big)^3.
\end{eqnarray*}
\end{rem}
}}

{\textcolor{blue}{
{\bf{Proof of Lemma 3.2.}} \\
Here, we give a proof only for the random variable $\Re(<J^{\gamma}_{P+} f_{\mathbb C}, \phi>)$ 
on the probability space  $({\cal S}'({\mathbb R}^d \to {\mathbb R}), {\cal B}({\cal S}'({\mathbb R}^d \to {\mathbb R}), \mu)$, since the other cases are similar ((3.19) follows from the arguments 
for $\Re(<J^{\gamma}_{P+} f_{\mathbb C}, \phi>)$, $\Im(<J^{\gamma}_{P+} f_{\mathbb C}, \phi>)$ and the analyticity of $\cos z$, $z \in {\mathbb C}$).  First of all we recall   
the description on the {\it{support property}} of $\mu_{Levy}$, which is given just after (2.3). 
Then, the probability measure $\mu_{Levy}$ has a support which is a sub {\it{Hilbert space}} of ${\cal S}'({\mathbb R}^d \to {\mathbb R})$, and 
 for {\it{real valued}} regular test functions $g \in {\cal S}({\mathbb R}^d \to {\mathbb R})$, 
 $<g, \phi>$, 
the dualization between $g$ and $\phi$, which 
 can be looked upon  as a {\it{stochastic integral}} 
of $g$ with respect to the {\it{Levy random field}} defined by (2.3), 
 is well-defined.
  By Lemma 3.1, we then  recall that 
  for $f_{\mathbb C} \in {\cal S}({\mathbb R}^d \to {\mathbb C})$, the functions 
  $J^{\gamma +}_{P+} f_{\mathbb C}$ and $J^{\gamma -}_{P+} f_{\mathbb C}$ are $L^p$   functions, $p \in [2,\infty]$, but they may not possess the sufficient regularities under which the dualizations between them and $\phi$ can be defined. However, by their integrability given by  (3.9), $<J_{P+}^{\gamma +} f_{\mathbb C}, \phi>$,  
$<J_{P+}^{\gamma -} f_{\mathbb C}, \phi>$ are  defined as $L_{\mu}^p$ 
 random variables, $\forall p \in [2, \infty)$,  in a sense of {\it{stochastic extension}}, in the framework of {\it{stochastic integeals}}. 
 Actually, for each $f_{\mathbb C} \in {\cal S}({\mathbb R}^d \to {\mathbb C})$, 
  we see that there exists a sequence $\{v_n \}_{n \in {\mathbb N}}$, 
  $v_n \in {\cal S}({\mathbb R}^d \to {\mathbb C})$, $n \in {\mathbb N}$ 
  (recall that on the probability space 
   $({\cal S}'({\mathbb R}^d \to {\mathbb R}), {\cal B}({\cal S}'({\mathbb R}^d \to {\mathbb R}), \mu_{Levy})$, the dualization $<v_n, \phi>$ is well-defined as an $L^p_{\mu_{Levy}}$ random variable, $p \in [2, \infty)$ by (2.3) with an obvious interpretation to the {\it{complex valued}} $v_n$), 
   and a random variable 
   \begin{equation}
   v(f_{\mathbb C}) \in  \bigcap_{p=2}^{\infty} L^p_{\mu},
   \end{equation}
  so that 
  \begin{equation}
  \lim_{n \to \infty}  \| <v_n, \phi> - v(f_{\mathbb C}) \|_{L^p_{\mu}} = 0,  
  \qquad \forall p \in [2, \infty),
  \end{equation}
  where the {\it{complex valued}}  random variable $v(f_{\mathbb C})$ satisfies 
  the equality (3.19) of which right hand side $\displaystyle \int_{{\cal S}'({\mathbb R}^d \to {\mathbb R})}
e^{i < J^{\gamma}_{P+} f_{\mathbb C}, \phi>} d \mu(\phi) $ is replaced by 
$\displaystyle \int_{{\cal S}'({\mathbb R}^d \to {\mathbb R})}
e^{i v(f_{\mathbb C}) }d \mu(\phi) $. 
  Then, $v(f_{\mathbb C})$ 
  is a {\it{stochastic extension}} of the dualization between 
  ${\cal S}({\mathbb R}^d \to {\mathbb C})$ and ${\cal S}'({\mathbb R}^d \to {\mathbb C})$,  which 
  defines the random variable $<J_{P+}^{\gamma +} f_{\mathbb C}, \phi>$ (the corresponding discussion for $<J_{P+}^{\gamma -} f_{\mathbb C}, \phi>$ 
  is completely analogous, and below we do not repeat it):
  \begin{equation}
  <J_{P+}^{\gamma +} f_{\mathbb C}, \phi>
 \stackrel{\mathrm{def}}{=} 
  v(f_{\mathbb C}).
  \end{equation}
}}

{\textcolor{blue}{
For each $J^{\gamma +}_{P+} f_{\mathbb C}$ with given $f_{\mathbb C} \in {\cal S}({\mathbb R}^d \to {\mathbb C})$, 
the construction of $\{ v_n \}_{n \in {\mathbb N}}$, by which (3.20) with (3.21) hold, 
can be carried out by 
  a standard argument on the theory of {\it{Lebesgue integrals}} and {\it{mollifier}} 
  (regularization).  {\it{A crucial  point}} on the statement (3.21) is that {\it{by one sequence}} $\{ v_n \}_{n \in {\mathbb N}}$ we have the convergence (3.21) for every $p \in [2, \infty)$. 
  Since the present standard argument is technical, we show it a little in detail step by step as follows: \\
  The {\it{complex valued}} function $J^{\gamma +}_{P+} f_{\mathbb C}$ is 
  precisely expressed by 
  $$J^{\gamma +}_{P+} f_{\mathbb C} = \Re J^{\gamma +}_{P+} f_{\mathbb C} + i \Im J^{\gamma +}_{P+} f_{\mathbb C},$$
with the decompositions 
  $$\Re J^{\gamma +}_{P+} f_{\mathbb C} = \big(\Re J^{\gamma +}_{P+} f_{\mathbb C} \big)_{+} - \big(\Re J^{\gamma +}_{P+} f_{\mathbb C} \big)_{-},
  \qquad 
  \Im J^{\gamma +}_{P+} f_{\mathbb C} = \big(\Im J^{\gamma +}_{P+} f_{\mathbb C} \big)_{+} - \big(\Im J^{\gamma +}_{P+} f_{\mathbb C} \big)_{-},
  $$
  where,  for a  {\it{real valued}} function $G(x)$, $x \in {\mathbb R}^d$, we denote 
  $(G(x))_{+} = \max \{G(x), 0\}$, $(G(x))_{-} = \max \{- G(x), 0\}$.
  Since, 
  on the consideration of 
   (3.20), (3.21) and (3.22), 
    for every above  decomposed term 
   the argument corresponding to $\big(\Re J^{\gamma +}_{P+} f_{\mathbb C} \big)_{+}$ is common, we give 
   it only for  
   $\big(\Re J^{\gamma +}_{P+} f_{\mathbb C} \big)_{+}$. \\
 i) \quad For each given $f_{\mathbb C} \in {\cal S}({\mathbb R}^d \to {\mathbb C})$, we {\it{firstly}} define 
 a sequence $\{u(1,n)\}_{n \in {\mathbb N}}$ such that 
\begin{equation}
u_{1,n}(x) = \left\{
  \begin{array}{ll}
     \displaystyle{ \frac{k-1}{2^n}}, & \quad x \in \{y \, : \, \frac{k-1}{2^n} \leq 
       (\Re J^{\gamma +}_{P+} f_{\mathbb C})_{+}(y) \leq \frac{k}{2^n}, \, \, |y| < n \}, \, \, 
          1 \leq k \leq 2^n n; \\ [0.4cm]
          \displaystyle{ 0}, & \quad x \in \{y \, : \, (\Re J^{\gamma +}_{P+} f_{\mathbb C})_{+}(y) \geq n\} \cup \{y \, : \, |y| \geq n\}
          \end{array}  \right.
          \end{equation}
          Then, 
 for each $x \in {\mathbb R}^d$,   $u_{1,n}(x)$ converges to   $(\Re J^{\gamma +}_{P+} f_{\mathbb C})_{+}(x)$ pointwise as $n \to \infty$. {\footnote{{\textcolor{blue}{ Actually, we have a stronger result than (3.9), for $\Re J^{\gamma +}_{P+}f_{\mathbb C}$, 
 $\Im J^{\gamma +}_{P+}f_{\mathbb C}$, $\Re J^{\gamma -}_{P+}f_{\mathbb C}$, $\Im J^{\gamma -}_{P+}f_{\mathbb C}$, for e.g., 
 $\Re J^{\gamma +}_{P+}f_{\mathbb C} \in H^n$ for any $n \in {\mathbb N}$  ($H^n$ denotes the {\it{Sobolev space}} of complex valued functions on ${\mathbb R}^d$ of which derivatives in the distribution sense up to $n \in {\mathbb N}$ are all in $L^2({\mathbb R}^d \to {\mathbb C}$), for the functions corresponding to the operations  $\Im J^{\gamma +}$, $\Re J^{\gamma -}$ and $\Im J^{\gamma -}$ have the same property.  In fact 
 for $f_{\mathbb C} \in {\cal S}({\mathbb R}^d \to {\mathbb C})$, 
 for any $n \in {\mathbb N}$ there exists a constant $C'$ and 
 ${\hat{f}}_{\mathbb C}$ satisfies not only (3.11)  but also 
 $|\hat{f} (\tau, r)| \leq \max \{ \frac{C'}{|\tau|^n}, \, \frac{C'}{r^n} \}$, $\forall (\tau, \xi) 
\notin S_m$. By making use of this bound, we clearly see that (cf. (3.12) also), 
for any $n \in {\mathbb N}$, 
$\int_{{\mathbb R}^d} | (\tau^2 + r^2)^n j^{\gamma}_{P+} (\tau, r) \, \hat{f}(\tau, r)|^2 \,  r^{d-1}  d \tau d r < \infty$, which shows that $\Re J^{\gamma}_{P+}f_{\mathbb C} \in H^n$ for any $n \in {\mathbb N}$. Then by {\it{Sobolev's embedding theorem}} (cf. e.g., Section VI-7 of  [YosidaK{\^o}saku1980], or Section 2-3 of 
[Mizohata 1973])
), $\Re J^{\gamma}_{P+}f_{\mathbb C}$ can be understood as a {\it{bounded continuous function}} on ${\mathbb R}^d$.}}
}}
          Since 
          $\Re J^{\gamma +}_{P+} f_{\mathbb C} \in L^p({\mathbb R}^d \to {\mathbb C})$, 
          $p \in [2, \infty]$, by (3.9), then 
           by {\it{Lebesgue's}} bounded convergence theorem, the sequence of {\it{simple functions}} $\{u_{1,n} \}_{n \in {\mathbb N}}$  converges to 
        $(\Re J^{\gamma +}_{P+} f_{\mathbb C})_{+}$ for all $p \in [2, \infty)$. 
 We then define a sequence of numbers $\{N(1,n)\}_{n \mathbb N}$ such that,
 for each $n \in {\mathbb N}$,
 \begin{equation}
 N(1,n)    \stackrel{\mathrm{def}}{=}   
 \max_{p = 2k, \, k \leq n} 
  \big\{ K(p,n) = \min_{n'} \{
  n' \, : \, \|u_{1,n'} - (\Re J^{\gamma +}_{P+} f_{\mathbb C})_{+} \|_{L^p} \leq \frac{1}{n}  \} \big\}.
 \end{equation}
 By (3.24),   for each given $f_{\mathbb C} \in {\cal S}({\mathbb R}^d \to {\mathbb C})$, 
 we define a sequence $\{u(2,n)\}_{n \in {\mathbb N}}$ as follows:
 \begin{equation*}
 u_{2,n} \stackrel{\mathrm{def}}{=} u_{1, N(1,n)}.
 \end{equation*} 
 Then, by (3.24) we have 
 \begin{equation}
 \|u_{2,n} - (\Re J^{\gamma +}_{P+} f_{\mathbb C})_{+} \|_{L^p}, \qquad 
  \forall p=2m, \quad m \leq n.
 \end{equation}
 ii) \quad 
 Each simple function $u_{2,n}(x)$, $x \in {\mathbb R}^d$ defined above can be expressed by  
 \begin{equation}
 u_{2,n}(x) = \sum_{j=1}^K \alpha_j {\chi}_{E_j}(x), \quad {\mbox{for some 
 $K \in {\mathbb N}$,  $\alpha_j >0$ and $E_j \subset B_{N(1,n)}$}}, \, j=1, \dots, K, 
 \end{equation} 
 where 
 $B_{N(1,n)} = \{y : |y| < N(1,n), y \in {\mathbb R}^d \}$, an {\it{open ball}}, and 
 ${\chi}_A(x)$ is the indicator function on a set $A \subset {\mathbb R}^d$. 
 For each {\it{measurable set}} $E_j$ and any $n\in {\mathbb N}$, there exists a {\it{closed set}} $F_j$ and an {\it{open set}} $G_j$ such that (if necessary we can take $G_j = G'_j \cap 
 B_{N(1,n)}$ for some adequate {\it{open set}} $G'_j$)
 \begin{equation}
 F_j \subset E_j \subset G_j, \subset B_{N(1,n)},  \qquad \lambda(G_j \setminus F_j) < 
 (\frac{1}{K n \alpha'_j})^{2n},
 \end{equation}
 where $\lambda(\cdot)$ is the {\it{Lebesgue measure}} on ${\mathbb R}^d$ and $\alpha'_j = \max \{1, \alpha_j \}$. 
 Next, as a standard procedure, for each $j =1, \dots, K$, take a {\it{continuous function}}  
 $h_j(x)$ on ${\mathbb R}^d$ such that $0 \leq h_j(x) \leq 1$ and 
 \begin{equation}
 h_j(x) = 1 \quad {\mbox{on $E_j$}}, \qquad h_j(x) = 0 \quad {\mbox{on ${\mathbb R}^d \setminus G_j$}},
 \end{equation}
  and then define 
 \begin{equation}
 u_{3,n}(x) \stackrel{\mathrm{def}}{=} \sum_{j=1}^K\alpha_j h_j(x).
 \end{equation}
 Then, by (3.26), (3.27), (3.28) and (3.29), 
  for $u_{3,n}$, a {\it{continuous function}} with {\it{compact support}},  
 we see that the following evaluation holds:
 \begin{equation}
 \|u_{2,n} - u_{3,n}\|_{L^p} \leq \sum_{j=1}^K |\alpha_j| \|{\chi}_{E_j} - h_j \|_{L^p} 
 < \sum_{j=1}^K \alpha_j  (\frac{1}{K n \alpha'_j})^{\frac{2n}{p}} \leq \frac{1}n, \quad 
  \forall p=2m, \, m \leq n.
 \end{equation}
 iii) \quad 
 Thirdly, 
  by applying the {\it{mollifier}} to $\{u_{3,n} \}_{n \in {\mathbb N}}$, defined by (3.29) for 
 for each given $f_{\mathbb C} \in {\cal S}({\mathbb R}^d \to {\mathbb C})$, 
 we construct 
 a sequence $\{v_n \}_{n \in {\mathbb N}}$, 
  $v_n \in {\cal S}({\mathbb R}^d \to {\mathbb C})$, $n \in {\mathbb N}$ by which (3.20) and (3.21) are satisfied. The corresponding procedure is standard in the framework of {\it{functional analysis}}, but here we review it. Recall that we can take a {\it{non-negative (real valued)}} function $\eta \in C^{\infty}_0({\mathbb R}^d)$ such that $support[\eta] \subset \{x\, :\, |x| \leq 1\}$ and $\displaystyle {\int_{{\mathbb R}^d} \eta(x) dx =1}$. Then, for $\epsilon >0$, define $\displaystyle{ \eta_{\epsilon}(x) 
  = \frac{1}{\epsilon^d} \eta(\frac{x}{\epsilon}) }$, $x \in {\mathbb R}^d$, by which a 
   {\it{mollifier}} ({\it{regularization}}),  is defined through a {\it{convolution}} (cf. e.g., Section VI-3 of [YosidaK{\^o}saku1980]).
 Namely, for each $n \in {\mathbb N}$, we can take an $\epsilon(n) >0$ in order that 
 the function $v_n \in C_0^{\infty}({\mathbb R}^d)$ 
 satisfies 
 \begin{equation}
 support[v_n] \subset B_{N(1,n) + \epsilon(n)}, \quad 
 \sup_{x \in {\mathbb R}^d} |v_n(x) - u_{3,n}(x)| \leq \frac{1}{n \lambda(B_{N(1,n) + \epsilon(n)})}, 
 \end{equation}
 where 
 \begin{equation}
 v_n(x)  \stackrel{\mathrm{def}}{=}  
 \int_{{\mathbb R}^d} \eta_{\epsilon(n)}(x-y) u_{3,n}(y) dy, \quad x \in {\mathbb R}^d,
 \end{equation}
 $B_{N(1,n) + \epsilon(n)} = \{y : |y| < N(1,n) + \epsilon(n), y \in {\mathbb R}^d \}$ 
 (see (3.27), (3.28), (3.29)) 
 and $\lambda(\cdot)$ is the {\it{Lebesgue measure}} on ${\mathbb R}^d$. 
 Define a sequence $\{v_n \}_{n \in {\mathbb N}}$, then by (3.31) we have 
 \begin{equation}
 \| v_n - u_{3,n} \|_{L^p({\mathbb R}^d)} \leq \frac{1}{n}, 
 \qquad
 \forall p=2m, \quad  m \leq n.
 \end{equation}
 iv) \quad Finally,
  for each given $f_{\mathbb C} \in {\cal S}({\mathbb R}^d \to {\mathbb C})$, 
  by (3.25), (3.30) and (3.33) all together, for any $n$ we have 
  \begin{equation}
  \|v_n - (\Re J^{\gamma +}_{P+} f_{\mathbb C})_{+} \|_{L^p({\mathbb R}^d)} 
  \leq \|v_n - u_{3,n}  \|_{L^p({\mathbb R}^d)} 
  + \|u_{3,n} - u_{2,n} \|_{L^p({\mathbb R}^d)} 
  + \|u_{2,n} - (\Re J^{\gamma +}_{P+} f_{\mathbb C})_{+} \|_{L^p({\mathbb R}^d)} 
  \leq \frac{3}{n}, 
  \end{equation}
 $$
 \forall p=2m, \quad  m \leq n.
 $$ 
 Recall that for $v_n \in C_0^{\infty}({\mathbb R}^d \to {\mathbb R}) \subset 
 {\cal S}({\mathbb R}^d \to {\mathbb R})$, the {\it{characteristic function}} (2.3), by replacing $g$ by $v_n$, holds, and the moments (expectations) $\int_{S'} (<v_n, \phi>)^{2m} d \mu(\phi) $ can be calculated through (2.3) (cf. (3.14) and (3.15)). In fact these moments are given, e.g., by the same formulas in  {\it{Remark 3.7}} by replacing $\Re (<J^{\gamma}_{P+} f_{\mathbb C}, \phi>)$ and $F$ by $<v_n, \phi>$ and $v_n$ respectively. This tells us that $\int_{S'} (<v_n, \phi>)^{2m} d \mu(\phi)$ is a {\it{polynomial}} of $v_n$ up to the power $2m$. Then, by (3.34), for each fixed $p = 2m$, we see that $\{<v_n, \phi>\}_{n \in {\mathbb N}}$ forms a {\it{Cauchy sequence}} in $L^p_{\mu}({\cal S}')$, and there exists a limit $\Re v(f_{\mathbb C})$ 
 in $L^p_{\mu}({\cal S}')$: 
 \begin{equation}
 \lim_{n \to \infty} <v_n, \phi> = \Re v(f_{\mathbb C}), \quad {\mbox{in $L^p_{\mu}({\cal S}')$}}.
 \end{equation} 
 By (3.34), since,
  by the same $\Re v(f_{\mathbb C})$, 
   the convergence (3.35) holds for all $p = 2m$, $m \in {\mathbb N}$, 
 we are able to define 
 \begin{equation}
  <(\Re J_{P+}^{\gamma +} f_{\mathbb C})_{+}, \phi>
 \stackrel{\mathrm{def}}{=} 
 \Re  v(f_{\mathbb C}) \in \bigcap_{p=2}^{\infty} L^p_{\mu}.
  \end{equation}
  Since, he corresponding consideration for  
  $<(\Re J_{P+}^{\gamma +} f_{\mathbb C})_{-}, \phi>$,  
  $<(\Im J_{P+}^{\gamma +} f_{\mathbb C})_{+}, \phi>$ 
  and
  $<(\Im J_{P+}^{\gamma +} f_{\mathbb C})_{-}, \phi>$ 
  are completely analogous to the above, $<J_{P+}^{\gamma +} f_{\mathbb C}, \phi>
  =
  <\Re J_{P+}^{\gamma +} f_{\mathbb C}, \phi>
 + i<\Im J_{P+}^{\gamma +} f_{\mathbb C}, \phi>$
 can be defined (see (3.36)).
 We have proven (3.20) and (3.21) (cf. (3.22)).
  }}

{\textcolor{blue}{
\begin{lmm}[Technical Lemma 3, ongoing]
Some technical arguments corresponding to 
 the property of spectrum of projection valued measure given by 
(2.18), (2.19) and (2.20) are put here.
\end{lmm}
}}

\section{Appendix}

\subsection{Foundations on Lorentz group} 
Suppose that $d=4$ and denote
$$x = (x_0, x_1, x_2, x_3),  \quad 
{\tilde{x}} = (x_0, -x_1, -x_2, -x_3) \in {\mathbb R}^4,$$
$$y = (y_0, y_1, y_2, y_3), \quad 
{\tilde{y}} = (y_0, -y_1, -y_2, -y_3) \in {\mathbb R}^4.
$$
The {\it{Lorentz scalar product}} between $x, \, y \in {\mathbb R}^4$ is defined by 
\begin{equation}
x \cdot {\tilde{y}} = x_0 \cdot y_0 - (x_1 \cdot y_1 + x_2 \cdot y_2 + x_3 \cdot y_3),
 \end{equation}
 The {\it{Lorentz group}} ${\cal L}$ is the set of linear transformations on ${\mathbb R}^4$ that preserve the {\it{Lorentz scalar product}} defined by (4.1). For ${\Lambda} \in {\cal L}$, we identify $\Lambda$ as the corresponding {\it{real}} $4 \times 4$ {\it{matrix}}, and use the same 
 notation $\Lambda$ for the matrix.
Then, by (4.1), $\Lambda \in {\cal L}$ is equivalent with 
\begin{equation}
({}^t \Lambda) \cdot  {\tilde E}\cdot  \Lambda = {\tilde E},
\end{equation}
where
\begin{equation}
\tilde{E} = \left(
\begin{array}{cccc}
1 &  0  &  0  &  0\\
0 & -1 &  0  &  0\\
0 &  0  & -1 &  0\\
0 &  0  &  0  &-1
\end{array}
\right).
\end{equation}
Then, by (4,2), for any $\Lambda \in {\cal L}$, 
\begin{equation*}
-1 = | {\tilde{E}}|  = | ({}^t \Lambda) \cdot  {\tilde E}\cdot  \Lambda| = - |\Lambda|^2,
\end{equation*}
thus
\begin{equation}
|\Lambda | = +1 \, \, {\mbox{or}} \, \, -1.
\end{equation}

The fundamental fact that $\cal L$ forms a {\it{group}}, i.e., 
for $\Lambda, \Gamma \in {\cal L}$ 
it satisfies 
\begin{equation}
\Lambda \cdot \Gamma \in {\cal L}, 
\end{equation}
\begin{equation}
{\Lambda}^{-1} \in {\cal L}, 
\end{equation}
and  an extra property
\begin{equation}
{}^t {\Lambda} \in {\cal L}, 
\end{equation}
can be easily certified as follows:\\
for $\Lambda, \Gamma \in {\cal L}$, since $\Lambda$ and $\Gamma$ satisfy (4.2),  it holds that 
\begin{eqnarray}
  {}^t(\Lambda \cdot \Gamma) {\tilde{E}} (\Lambda \cdot \Gamma)  & = & 
  {}^t \Gamma \cdot {}^t \Lambda {\tilde{E}}  \Lambda \cdot \Gamma \nonumber \\
     & = & {}^t \Gamma {\tilde{E}} \Gamma = {\tilde{E}}.
     \end{eqnarray}
By (4.2), (4.8) is equivalent with (4.5). \\
Next, by (4.4) since $\Lambda \in {\cal L}$ is nonsingular and 
$${}^t( \Lambda^{-1}) = ({}^t \Lambda)^{-1},$$
by (4.2) it holds that 
\begin{eqnarray}
  {\tilde{E}} & = &
   E \cdot {\tilde{E}} \cdot E = ({}^t(\Lambda^{-1}) \cdot {}^t \Lambda) 
  {\tilde{E}} (\Lambda \cdot \Lambda^{-1}) \nonumber \\
     & = & {}^t(\Lambda^{-1}) \cdot ( {}^t \Lambda {\tilde{E}} \Lambda) \cdot {\Lambda}^{-1} 
     = {}^t (\Lambda^{-1}) \cdot {\tilde{E}} \cdot (\Lambda^{-1}).
     \end{eqnarray}
Again, by (4.2), (4.8) is equivalent with (4.6).\\
Finally, by (4.2), since ${\tilde{E}}^{-1} = {\tilde{E}}$, it holds that
\begin{equation}
\Lambda {\tilde{E}} ({}^t \Lambda) {\tilde{E}} = {\tilde{E}} ({}^t \Lambda) {\tilde{E}} \Lambda \Lambda^{-1} = \Lambda  {\tilde{E}} {\tilde{E}} \Lambda^{-1} = E.
\end{equation}
Again, by making use of ${\tilde{E}}^{-1} = {\tilde{E}}$, from (4.14) it holds that 
\begin{equation}
\Lambda \cdot {\tilde{E}} \cdot ( {}^t \Lambda) = E {\tilde{E}} = {\tilde{E}}.
\end{equation}
By (4.2), (4.11) is equivalent with (4.7).

{\textcolor{red}{
For example, these matrices such that
}}
\begin{equation}
{\Lambda}_1(\theta) = \left(
\begin{array}{cccc}
1 &  0  &  0  &  0\\
0 & \cos (\theta) &  \sin (\theta)  &  0\\
0 &  - \sin (\theta)  & \cos (\theta) &  0\\
0 &  0  &  0  &1
\end{array}
\right) 
\qquad {\mbox{with a  variable $\theta \in {\mathbb R}$}}, 
\end{equation}
\begin{equation}
{\Lambda}_2(\eta) = \left(
\begin{array}{cccc}
\cosh (\eta) &  0  &  0  &  \sinh (\eta)\\
0 & 1 &  0  &  0\\
0 &  0  & 1 &  0\\
\sinh (\eta) &  0  &  0  & \cosh (\eta)
\end{array}
\right) \qquad 
{\mbox{with a  variable $\eta \in {\mathbb R}$}},
\end{equation}
\begin{equation}
{\Lambda}_3(\theta, \eta) = {\Lambda}_1(\theta) \cdot {\Lambda}_2(\eta) 
= \left(
\begin{array}{cccc}
\cosh (\eta) &  0  &  0  &  \sinh (\eta)\\
0 & \cos (\theta) &  \sin (\theta)  &  0\\
0 &  - \sin (\theta)  & \cos (\theta) &  0\\
\sinh (\eta) &  0  &  0  & \cosh (\eta)
\end{array}
\right), 
\end{equation}
\begin{equation}
{\Lambda}_4(\beta) = \left(
\begin{array}{cccc}
1+ \frac12 |\beta|^2 &  \Re (\beta)  &  - \Im (\beta)  &  - \frac12 |\beta|^2 \\
\Re (\beta) & 1 &  0  &  - \Re (\beta)\\
- \Im (\beta) &  0  & 1 &  \Im (\beta)\\
\frac12 |\beta|^2 &  \Re (\beta)  &  - \Im (\beta)  & 1- \frac12 |\beta|^2
\end{array}
\right)
\qquad {\mbox{with  a  variable $\beta  \in {\mathbb C}$}}, 
\end{equation}
and
\begin{equation*}
 \left(
\begin{array}{cccc}
1 &  0  &  0  &  0\\
0 & 1 &  0  &  0\\
0 &  0  & 1 &  0\\
0 &  0  &  0  &1
\end{array}
\right), 
\quad 
\left(
\begin{array}{cccc}
-1 &  0  &  0  &  0\\
0 & -1 &  0  &  0\\
0 &  0  & 1 &  0\\
0 &  0  &  0  &1
\end{array}
\right),
\quad 
\left(
\begin{array}{cccc}
1 &  0  &  0  &  0\\
0 & -1 &  0  &  0\\
0 &  0  & -1 &  0\\
0 &  0  &  0  &1
\end{array}
\right),
\quad 
\left(
\begin{array}{cccc}
1 &  0  &  0  &  0\\
0 & 1 &  0  &  0\\
0 &  0  & -1 &  0\\
0 &  0  &  0  &-1
\end{array}
\right),
\end{equation*}
are elements in ${\cal L}$ by (4.2), and their determinant are $|\Lambda| = 1$, 
correspondingly, the matrices such that 
\begin{equation*}
\tilde{E} = \left(
\begin{array}{cccc}
1 &  0  &  0  &  0\\
0 & -1 &  0  &  0\\
0 &  0  & -1 &  0\\
0 &  0  &  0  &-1
\end{array}
\right), \qquad 
\left(
\begin{array}{cccc}
-1 &  0  &  0  &  0\\
0 & 1 &  0  &  0\\
0 &  0  & 1 &  0\\
0 &  0  &  0  &1
\end{array}
\right),
\qquad 
\left(
\begin{array}{cccc}
1 &  0  &  0  &  0\\
0 & 0 &  0  &  1\\
0 &  0  & 1 &  0\\
0 &  1  &  0  &0
\end{array}
\right),
\end{equation*}
are also elements in  ${\cal L}$, but  their determinant are $|\Lambda| = -1$.

{\textcolor{red}{
The subgroup of all Lorentz transformations preserving both orientation and direction of 
 the first component of $x \in {\mathbb R}^4$, i.e., $x_0$ (or the time component), hence  satisfying  both 
 for the $(0,0)$ component 
 of the $d\times d$ matrix $\Lambda$, 
 $\Lambda_{0,0} >0$, 
 and
$| \Lambda | = +1$ (see (4.4)), is called as the {\it{proper, orthochronous Lorentz group}} or 
{\it{restricted Lorentz group}}.
 In other words, the restricted Lorentz group is the 
 largest subgroup connected to 
the identity $E$.
}}
As the {\it{Lie group}} notation (cf. e.g., [Wikipedia]) the subgroup is denoted by $SO^+(1,3)$, and here we denote it by $\Lambda \in {\cal L}_+^{\uparrow}$ (see section 1). Hence, the conjugacy classes of the {\it{restricted Lorentz group}} fall into following five classes: The identity transformations; Elliptic transformations; Hyperbolic transformations; Loxodromic transformations; Parabolic transformations. 
{\textcolor{red}{
 These matrices 
$E$, the identity, and 
$\Lambda_1(\theta)$, $\Lambda_2(\eta)$, $\Lambda_3(\theta, \eta)$, $\Lambda_4(\beta)$  given by (4.12), (4.13), (4.14), (4.15), respectively,  correspond with above five classes respectively 
}}
(cf. e.g. [Wikipedia] of the English version, there these matrices are denoted by $Q_1$, $Q_2$, $Q_3$ and $Q_4$ respectively (in the Japanese version of the corresponding [Wikipedia] the formula $Q_4$ seems incorrect, as far as the version before 
 May 2024)).


\end{document}